\documentclass[reprint,amsmath,showpacs,amssymb,aip]{revtex4-1}

\usepackage[english]{babel}
\usepackage{graphicx}% Include figure files
\usepackage{dcolumn}% Align table columns on decimal point
\usepackage{bm}% bold math%
\usepackage{hyperref}

\newcommand{\figdir}{}

\begin{document}

\title{Transport and optical properties of warm dense aluminum in the two-temperature regime: \textit{Ab initio} calculation and semiempirical approximation}

\author{D.~V.~Knyazev$^{1,2,3}$ and P.~R.~Levashov$^{1,4}$}
\affiliation{$^{1}$Joint Institute for High Temperatures RAS, Izhorskaya 13 bldg. 2, Moscow 125412, Russia \\ 
$^2$Moscow Institute of Physics and Technology (State University), Institutskiy per. 9, Dolgoprudny, Moscow Region, 141700, Russia,
$^3$State Scientific Center of the Russian Federation --– Institute for Theoretical and Experimental Physics of National Research Centre "Kurchatov Institute", Bolshaya Cheremushkinskaya 25, 117218, Moscow, Russia \\
$^4$Tomsk State University, Lenin Prospekt 36, Tomsk, 634050, Russia}

\begin{abstract}
This work is devoted to the investigation of transport and optical properties of liquid aluminum in the two-temperature case. At first optical properties, static electrical and thermal conductivities were obtained in the \textit{ab initio} calculation. The \textit{ab initio} calculation is based on the quantum molecular dynamics, density functional theory and the Kubo-Greenwood formula. The semiempirical approximation was constructed based on the results of the \textit{ab initio} caculation. The approximation yields the dependences $\sigma_{1_\mathrm{DC}}\propto1/T_i^{0.25}$ and $K\propto T_e/T_i^{0.25}$ for the static electrical conductivity and thermal conductivity, respectively. The approximation is valid for liquid aluminum at $\rho=2.70$~g/cm$^3$, 3~kK~$\leq T_i\leq T_e\leq20$~kK. Our results are well described by the Drude model with the effective relaxation time $\tau\propto T_i^{-0.25}$. We have compared our results with a number of other models. They are all reduced in the low-temperature limit to the Drude model with different expressions for the relaxation time $\tau$. Our results are not consistent with the models in which  $\tau\propto T_i^{-1}$ and support the models which use the expressions with the slower decrease of the relaxation time.
\end{abstract}

\maketitle

\section{INTRODUCTION}

The experiments on the interaction of femtosecond lasers with matter are widespread nowadays \cite{Chen:PRL:2013, Widmann:PRL:2004, Chimier:PRB:2011, Veysman:JPB:2008}. During the ultrafast heating the matter passes through the two-temperature state with hot electrons and relatively cold ions. Different approaches may be used to simulate these experiments. Two-temperature hydrodynamic model \cite{Povarnitsyn:ASS:2012} and combined continuum-atomistic modeling \cite{Ivanov:PRB:2003} are widely used.

The information on the properties of matter is necessary for the practical implementation of the computational schemes. The necessary properties may differ depending on the type of simulation.

For thin targets at initial stage only the information on optical properties, electron and ion heat capacities, and electron-ion coupling is necessary \cite{Chen:PRL:2013}. The study of the optical properties is possible both using modern picosecond diagnostics \cite{Chen:PRL:2013, Widmann:PRL:2004} and computational methods \cite{Mazevet:PRL:2005,Chen:PRL:2013}.

Thermal conductivity is necessary, in addition, for the simulation of thick targets \cite{Kirkwood:PRB:2009, Loboda:HEDP:2011}. During the initial stage of experiment the density of the target remains almost constant, so the study of the matter properties along the normal isochor is of the paramount importance.

If the whole process of the target heating and post-evolution is to be studied, the full hydrodynamic simulation is necessary \cite{Povarnitsyn:ASS:2012}. Two-temperature equation of state, valid in the broad region of densities and temperatures, should be available.

The properties of matter in the two-temperature regime differ significantly from those in the equilibrium case. This has been investigated previously for structural properties \cite{Dharma-wardana:PRE:2008}, optical properties \cite{Mazevet:PRL:2005}, melting temperatures \cite{Recoules:PRL:2006, Ernstorfer:Science:2009}, density of states \cite{Lin:PRB:2008, Stegailov:CPP:2010}, phonon spectra \cite{Stegailov:CPP:2010}, electron-ion coupling factor \cite{Dharma-wardana:PRE:1998} and equations of state \cite{Levashov:JPCM:2010, Sinko:HEDP:2013}.

In this work we focus on the investigation of the optical properties, static electrical and thermal conductivities of liquid aluminum at normal density in the two-temperature case.

The usual approach to obtain transport and optical properties in the wide range of parameters is to interpolate between high- and low-temperature asymptotics \cite{LeeMore:PhysFluids:1984,Eidmann:PRE:2000,Povarnitsyn:ASS:2012}.

At high temperatures the electrons are non-degenerate and the ions are not correlated. The Spitzer theory \cite{Spitzer:PR:1953} is valid for the non-degenerate fully ionized plasma with a small non-ideality. At lower temperatures coupling, degeneracy corrections \cite{LeeMore:PhysFluids:1984} and partial ionization \cite{Apfelbaum:CPP:2013} should be taken into account. 

At low temperatures the electrons become degenerate and should be treated within the quantum mechanical approach. The optical properties may be obtained using the average atom models \cite{Johnson:JQSRP:2006,Starrett:PhysPlasmas:2012}. However, at low temperatures ion-ion correlations become important, and the coherent scattering of electrons by the ions should be taken into account. The average atom models do not allow for ion-ion scattering and do not give correct static electrical and thermal conductivities \cite{Johnson:HEDP:2009}.

The correct treatment of ion-ion correlations and coherent electron scattering is crucial for the calculations at relatively low temperatures. There are several approaches to deal with the problem.

The simplest approach is to use the Drude model with the effective collision frequency of the crystalline-like form $\nu_\mathrm{eff}=AT_i+BT_e^2$ \cite{LeeMore:PhysFluids:1984,Ivanov:PRB:2003}.

In the more complicated approximations the temperature growth of the effective collision frequency in the liquid phase is limited by some maximum value \cite{Eidmann:PRE:2000,Povarnitsyn:ASS:2012}. This approach is often used in the simulations of femtosecond laser heating \cite{Kirkwood:PRB:2009,Loboda:HEDP:2011,Povarnitsyn:ASS:2012}.

The ionic structure factor may be calculated by solving the integral equations for distribution functions \cite{Apfelbaum:JPhysA:2006, Dharma-wardana:PRE:2006}. Transport and optical properties are then obtained via the Ziman theory \cite{Apfelbaum:JPhysA:2006, Dharma-wardana:PRE:2006}. However, 
the treatment of the band excitation in the metals with complex electronic structure is difficult within this formalism.

The method based on the quantum molecular dynamics (QMD), density functional theory (DFT) and the Kubo-Greenwood formula is widely used to treat the problem of ion-ion correlations. This method was started in papers \cite{Fois:PRB:1989,Silvestrelli:PRB:1996,Silvestrelli:PRB:1999} and followed by well-known works \cite{Collins:PRB:2001,Desjarlais:PRE:2002,Recoules:PRB:2005}. Since that time the method was applied to the wide variety of materials \cite{Laudernet:PRB:2004,French:PhysPlasmas:2011,Lambert:PhysPlasmas:2011,Norman:CPP:2013,Wang:PhysPlasmas:2014}. In the present paper we use this \textit{ab initio} method to calculate transport and optical properties of aluminum in equilibrium and non-equilibrium regimes. 

Some additional contribution due to electron-electron collisions may be added to the results of the \textit{ab initio} calculation \cite{Inogamov:JETP:2010}.

Our paper is organized as follows. At first we briefly describe the numerical method (Section~\ref{Sec:Technique}). The calculation parameters are listed in Section~\ref{Sec:Parameters}. The results of our work are presented in Section~\ref{Sec:Results}. At first we describe the results of our \textit{ab initio} calculations (Subsection~\ref{Subsec:Results:Abinitio}). Then we build the semiempirical approximation of the obtained \textit{ab initio} results (Subsection~\ref{Subsec:Results:Approximation}). Then our \textit{ab initio} results and the approximation constructed are compared with the models of other authors (Subsection~\ref{Subsec:Results:Models}).

\section{CALCULATION TECHNIQUE}
\label{Sec:Technique}

Here we will give the brief overview of the calculation method. The more detailed description is available in our previous work \cite{Knyazev:COMMAT:2013} and earlier papers \cite{Collins:PRB:2001,Desjarlais:PRE:2002,Recoules:PRB:2005,Holst:PRB:2011}.

The calculation consists of three main stages: QMD simulation, precise resolution of the band structure, and the calculation of transport and optical properties via the Kubo-Greenwood formula.

At the first stage the atoms at the given density are placed to the supercell with periodic boundary conditions. The QMD simulation is performed. At each ionic step the electronic structure is calculated in the framework of finite-temperature DFT. Electrons are treated within the Born-Oppenheimer approximation: at each step electrons totally adjust to the current ionic positions. Electron temperature $T_e$ is set by the parameter in the Fermi-Dirac distribution. The ions are treated classically. The Hellmann-Feynman forces are calculated to propagate ionic trajectories in time. The ion temperature $T_i$ is maintained via the Nos\'e-Hoover thermostat. Independent ionic configurations are selected from the equilibrium stage of the QMD simulation.

At the second stage the precise resolution of the band structure is performed for the selected ionic configurations. As well as during the QMD simulation the electronic structure is calculated. However, the larger energy cut-off, number of bands, number of \textbf{k}-points in the Brillouin zone are used. The higher values of these parameters increase the precision of calculation. During the precise resolution of the band structure the electronic wave functions and energy eigenvalues are obtained. Further this information is used to calculate transport and optical properties using the Kubo-Greenwood formula.

At the third stage the Onsager coefficients are calculated using the Kubo-Greenwood formula. The static Onsager coefficients $\mathcal{L}_{mn},~m,n=1,2$ connect the applied electric field $\mathbf{E}$ and the electron temperature gradient $\nabla T_e$ with the emerging electric $\mathbf{j}$ and heat $\mathbf{j}_q$ current densities:
\begin{eqnarray}
\mathbf{j}=\frac{1}{e}\left(e\mathcal{L}_{11}\mathbf{E}-\frac{\mathcal{L}_{12}\nabla T_e}{T_e}\right),
\label{Eq:OnsagerDefinition1}
\\
\mathbf{j}_q=\frac{1}{e^2}\left(e\mathcal{L}_{21}\mathbf{E}-\frac{\mathcal{L}_{22}\nabla T_e}{T_e}\right).
\label{Eq:OnsagerDefinition2}
\end{eqnarray}
Here $e$ is the electron charge. For the sake of convenience we also introduce the Onsager coefficients $L_{mn},~m,n=1,2$ which are connected with $\mathcal{L}_{mn}$ via the following relations:
\begin{eqnarray}
L_{11}=\mathcal{L}_{11},~L_{12}=\frac{\mathcal{L}_{12}}{eT_e},~L_{21}=\frac{\mathcal{L}_{21}}{e},~L_{22}=\frac{\mathcal{L}_{22}}{e^2T_e}.
\end{eqnarray}

The dynamic Onsager coefficients $\mathcal{L}_{mn}(\omega),~m,n=1,2$ are calculated according to the Kubo-Greenwood formula:
\begin{multline}
\mathcal{L}_{mn}(\omega)=(-1)^{m+n}\frac{2\pi e^2\hbar^2}{3m_e^2\omega\Omega}\times\\
\sum_{i,j,\alpha,\mathbf{k}}W(\mathbf{k})\left(\frac{\epsilon_{i,\mathbf{k}}+\epsilon_{j,\mathbf{k}}}{2}-\mu\right)^{m+n-2}\left|\left\langle\Psi_{i,\mathbf{k}}\left|\nabla_\alpha\right|\Psi_{j,\mathbf{k}}\right\rangle\right|^2\\\times
\left[f(\epsilon_{i,\mathbf{k}})-f(\epsilon_{j,\mathbf{k}})\right]\delta(\epsilon_{j,\mathbf{k}}-\epsilon_{i,\mathbf{k}}-\hbar\omega).
\label{Eq:DynamicOnsager}
\end{multline}
Here $\Psi_{i,\mathbf{k}}$ and $\epsilon_{i,\mathbf{k}}$ are the electronic wave function and energy eigenvalue respectively, corresponding to the particular band $i$ and point in the Brillouin zone $\mathbf{k}$. This information is obtained during the precise resolution of the band structure. $f(\epsilon_{i,\mathbf{k}})$ is the Fermi-weight of the particular band, $W(\mathbf{k})$---the weight of the particular \textbf{k}-point. $\mu$ is the chemical potential, $\omega$---the frequency of the applied electric field, $\Omega$---the volume of the supercell. $\hbar$ is the reduced Planck constant, $m_e$ is the electron mass.

The delta-function in the Kubo-Greenwood formula (\ref{Eq:DynamicOnsager}) is broadened by the Gaussian function \cite{Desjarlais:PRE:2002}. The intuitively clear derivation of the Kubo-Greenwood formula for $\sigma_1(\omega)=\mathcal{L}_{11}(\omega)$ may be found in the paper \cite{Moseley:AJP:1978}. The derivation of the Kubo-Greenwood formula in the form (\ref{Eq:DynamicOnsager}) was performed in the paper \cite{Holst:PRB:2011}. It is worth noting, that the energy eigenvalues half-sum $\frac{\epsilon_{i,\mathbf{k}}+\epsilon_{j,\mathbf{k}}}{2}$ is used during the calculation. This was first established in the paper \cite{Holst:PRB:2011}, some additional discussion is also present in our previous paper \cite{Knyazev:COMMAT:2013}.

The real part of the dynamic electrical conductivity is readily obtained as $\sigma_1(\omega)=\mathcal{L}_{11}(\omega)$. The imaginary part is reconstructed via the Kramers-Kronig relation. If the dynamic electrical conductivity is known, we may calculate other optical properties \cite{Collins:PRB:2001,Knyazev:COMMAT:2013}: complex dielectric function, complex refraction index, reflectivity, and absorption coefficient.

If we calculate the transport properties, the static Onsager coefficients $\mathcal{L}_{mn}$ and $L_{mn}$ are obtained by the simple linear extrapolation to the zero frequency \cite{Knyazev:COMMAT:2013}. Except for the $\mathcal{L}_{11}(\omega)=\sigma_1(\omega)$, no physical meaning is assigned to the dynamic Onsager coefficients. They are necessary only to calculate the static ones. Then the transport properties are expressed as follows:
\begin{eqnarray}
\sigma_{1_\mathrm{DC}}=L_{11},
\\
K=L_{22}-\frac{L_{12}L_{21}}{L_{11}}.
\label{Eq:K_definition}
\end{eqnarray}
Here $\sigma_{1_\mathrm{DC}}$ stands for the static electrical conductivity, whereas $K$---for electron thermal conductivity. The thermal conductivity (\ref{Eq:K_definition}) contains not only the $L_{22}$ Onsager coefficient. Additional positive contribution, called thermoelectric term, is also subtracted. At low temperatures the relative contribution of the thermoelectric term is small \cite{Ashcroft:1976}.

The experimentally discovered Wiedemann-Franz law exists \cite{Ashcroft:1976}:
\begin{equation}
\frac{K(T)}{\sigma_{1_\mathrm{DC}}(T)\cdot T}=\mathrm{const}=L=\frac{\pi^2}{3}\frac{k^2}{e^2}.
\label{Eq:WiedemannFranz}
\end{equation}
The static electrical conductivity $\sigma_{1_\mathrm{DC}}(T)$ and thermal conductivity $K(T)$ depend on the temperature, whereas the ratio $L$, called the Lorenz number, does not depend on the temperature. The value $L=\frac{\pi^2}{3}\frac{k^2}{e^2}=2.44\cdot 10^{-8}$~W$\cdot\Omega\cdot$K$^{-2}$ is called the ideal value. The Wiedemann-Franz law was discovered at low temperatures in the equilibrium case $T_i=T_e$. In this work we obtain the static electrical and thermal conductivities and check the validity of the Wiedemann-Franz law at high temperatures in the non-equilibrium case.

The QMD simulation and the band structure calculation are performed using Vienna \textit{Ab initio} Simulation Package (VASP) \cite{Kresse:PRB:1993,Kresse:PRB:1994,Kresse:PRB:1996}. The special parallel program module was created to compute the dynamic Onsager coefficients, extrapolate them to zero frequency and calculate transport properties. This module uses information from the VASP package.

\section{CALCULATION PARAMETERS}
\label{Sec:Parameters}

The optimal choice of technical parameters was described in detail in our previous paper \cite{Knyazev:COMMAT:2013}. Here we will only specify the values of technical parameters.

The QMD simulation is performed for the supercell, containing 256 atoms. Initially the atoms are placed to the nodes of the fcc lattice. The Newton's motion equations are integrated using the Verlet algorithm. The integration step is 2~fs, 1500 steps of the QMD simulation are performed. The ultrasoft pseudopotential (US-PP) of Vanderbilt \cite{Vanderbilt:PRB:1990} together with the local density approximation (LDA) for the exchange correlation functional are used during the QMD simulation. According to the paper \cite{Levashov:JPCM:2010} the influence of the inner electrons on the thermodynamic properties becomes significant only at $T_e\lesssim80$~kK. In this work $T_e\leq50$~kK, so only three outer electrons were taken into account. The electronic structure is calculated for 1 \textbf{k}-point in the Brillouin zone ($\Gamma$-point) and energy cut-off $E_\mathrm{cut}=100$~eV. The number of bands is different for different temperatures. The number of bands is selected as described in \cite{Knyazev:COMMAT:2013}. Every 100 steps the configuration (except for the initial configuration) is selected for the further calculation of the transport and optical properties.

During the precise resolution of the band structure the electronic structure is calculated with the same pair US-PP, LDA, as during the QMD simulation. The energy cut-off is increased to $E_\mathrm{cut}=200$~eV. The higher number of bands, chosen as shown in \cite{Knyazev:COMMAT:2013}, is used. As well as during the QMD simulation, $\Gamma$-point only is used.

The dynamic Onsager coefficients are calculated for the frequencies from 0.005~eV up to 10~eV with the step 0.005~eV. The optimal broadening of the $\delta$-function in the Kubo-Greenwood formula $\Delta E=0.1$~eV was found \cite{Knyazev:COMMAT:2013}. The static Onsager coefficients were calculated by the linear extrapolation of the dynamic coefficients to the zero frequency (see \cite{Knyazev:COMMAT:2013} for details).

All the ($\rho$,$T_i$,$T_e$) points in this work were calculated with the parameters specified above. The only exception are the points with $T_e=50$~kK. They were calculated with 108 atoms while all other parameters were kept the same.

The error of the static electrical conductivity calculation was estimated as 20\% for liquid aluminum at $\rho=2.249$~g/cm$^3$, $T=1273$~K in our previous paper \cite{Knyazev:COMMAT:2013} by variation of the technical parameters of the method.

Our previous works \cite{Knyazev:COMMAT:2013,Povarnitsyn:CPP:2012} contain the comparison of our calculation with the similar computations of other authors, experimental and reference data. The agreement of our data with other results was permanently good.

\section{RESULTS AND DISCUSSION}
\label{Sec:Results}

\subsection{The results of \textit{ab initio} calculation}\
\label{Subsec:Results:Abinitio}

The calculations were performed for aluminum at normal density. The temperatures of electrons and ions were in the range from 3~kK up to 50~kK. Three basic types of the calculations were performed: 1) the equilibrium case with $T_e=T_i$ varied from 3~kK up to 20~kK; 2) at fixed ion temperature $T_i=3$~kK with $T_e$ varied from 3~kK up to 20~kK; 3) at fixed electron temperature $T_e=20$~kK with $T_i$ varied from 3~kK up to 20~kK. The variation of only one temperature with another one kept fixed makes possible to examine the influence of each temperature on transport and optical properties.

\begin{figure*}
	a)\includegraphics[width=0.95\columnwidth]{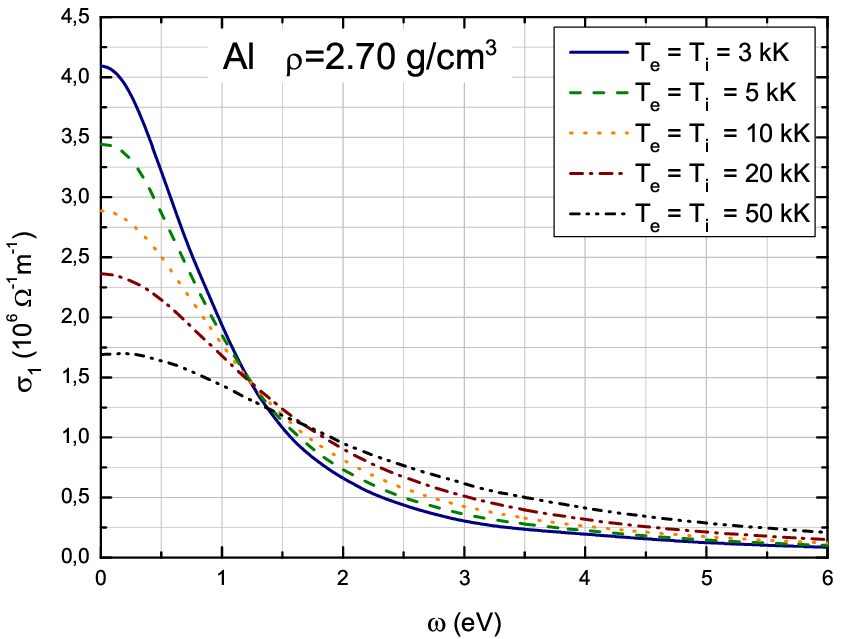}
	b)\includegraphics[width=0.95\columnwidth]{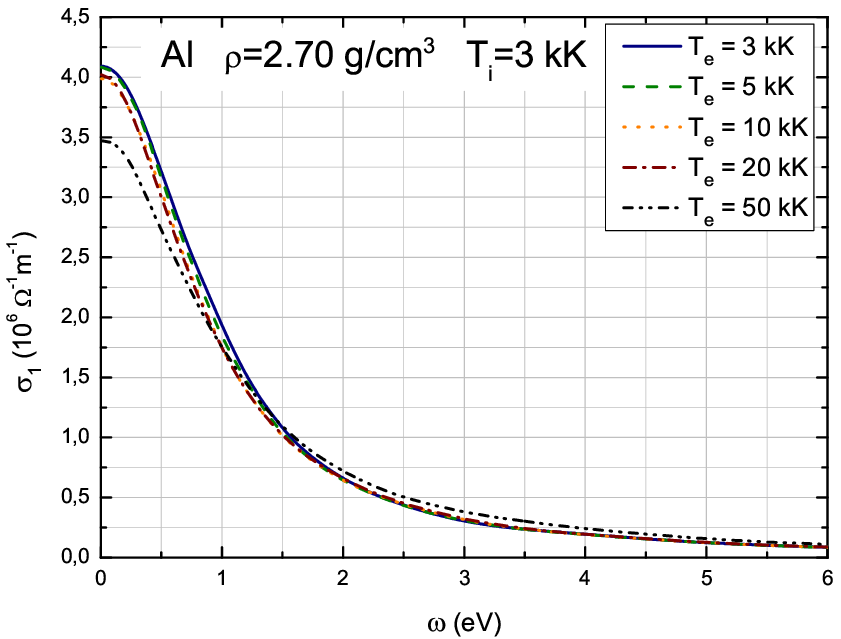}
	\caption{(Color online) The frequency dependences of the real part of dynamic electrical conductivity for different temperatures. a)~Equilibrium case $T_e=T_i$. b)~Non-equilibrium case, $T_i$ is fixed at 3~kK.}
	\label{Fig:Al_resigma}
\end{figure*}

The frequency dependences of the real part of dynamic electrical conductivity for different temperatures are shown in Fig.~\ref{Fig:Al_resigma}. All the curves have the Drude-like shapes. 

In the equilibrium case (Fig.~\ref{Fig:Al_resigma}a) the curves change significantly as the temperature is varied. At a fixed low frequency ($\omega \lesssim 1$~eV) the conductivity decreases
with temperature rise, at a high frequency ($\omega \gtrsim 2$~eV)---increases.

If $T_i$ is kept fixed at 3~kK, the curves for different $T_e$ almost coincide (Fig.~\ref{Fig:Al_resigma}b).

\begin{figure*}
	a)\includegraphics[width=0.95\columnwidth]{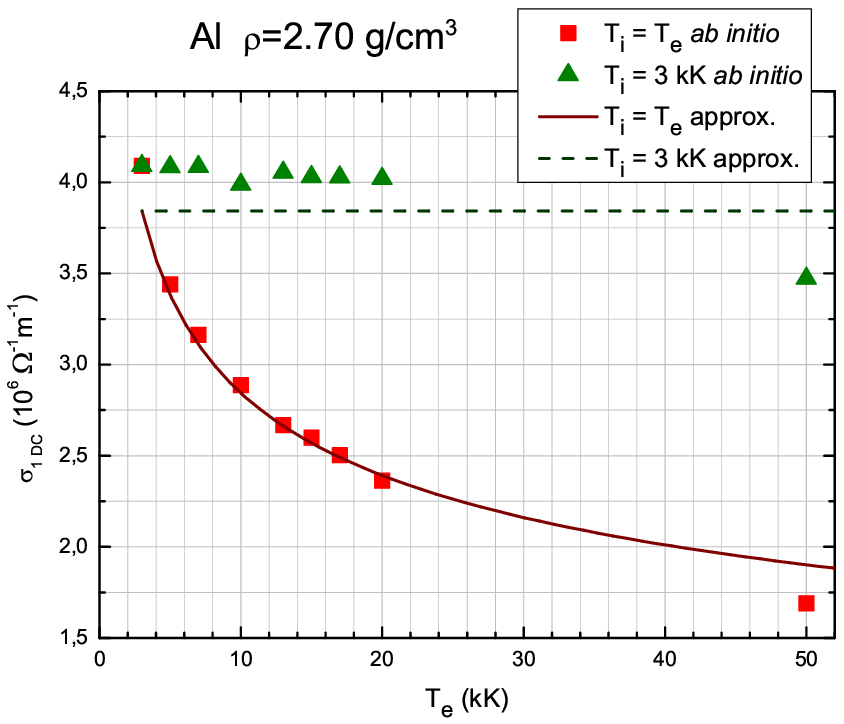}
	b)\includegraphics[width=0.95\columnwidth]{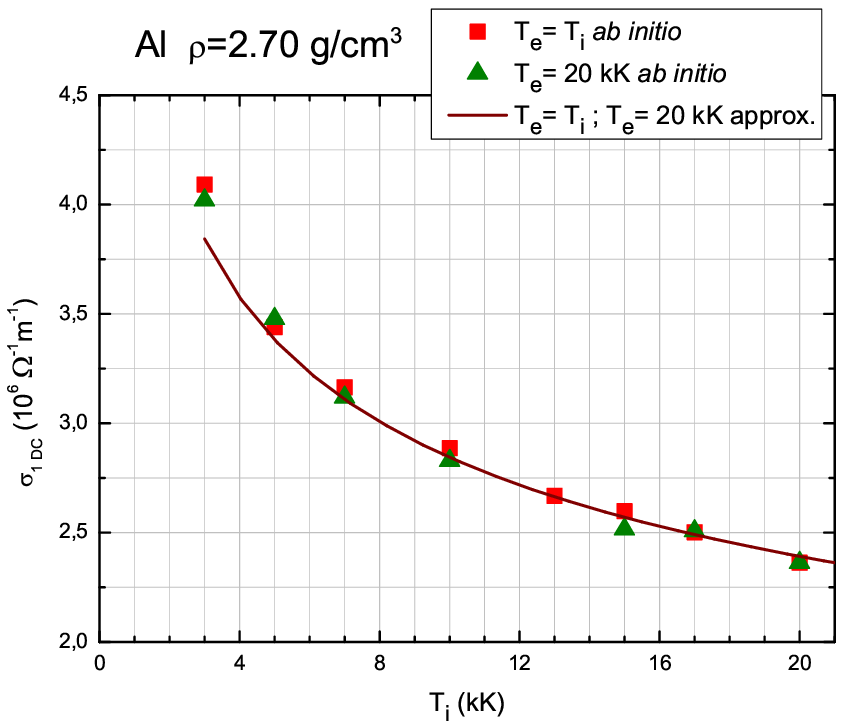}
	\caption{(Color online) Temperature dependences of static electrical conductivity. a) Dependences on the electron temperature $T_e$. Squares---equilibrium case $T_i=T_e$; triangles---non-equilibrium case, $T_i=3$~kK, $T_e$ is varied. Solid line---approximation (\ref{Eq:ResigmaDC_approx}) in equilibrium case, dashed line---approximation (\ref{Eq:ResigmaDC_approx}) at fixed $T_i=3$~kK. b) Dependences on the ion temperature $T_i$. Squares---equilibrium case $T_e=T_i$; triangles---non-equilibrium case, $T_e=20$~kK, $T_i$ is varied. Solid line---approximation (\ref{Eq:ResigmaDC_approx}) both for equilibrium and non-equilibrium cases.}
	\label{Fig:Al_resigmaDC}
\end{figure*}

The temperature dependences of the static electrical conductivity are shown in Fig.~\ref{Fig:Al_resigmaDC}.

In Fig.~\ref{Fig:Al_resigmaDC}a the electron temperature $T_e$ is varied. If it is changed together with the ion temperature, $T_i=T_e$, the conductivity decreases. But if $T_i$ is kept fixed at 3~kK, the conductivity
changes not so drastically. For the temperatures from 3~kK up to 20~kK the conductivity remains almost constant.

In Fig.~\ref{Fig:Al_resigmaDC}b the ion temperature $T_i$ is varied. The equilibrium dependence $T_i=T_e$ is obviously the same as in Fig.~\ref{Fig:Al_resigmaDC}a. If $T_e$ is kept fixed at 20~kK, and only $T_i$ is varied, the results almost coincides with the equilibrium data (Fig.~\ref{Fig:Al_resigmaDC}b).

\begin{figure*}
	a)\includegraphics[width=0.95\columnwidth]{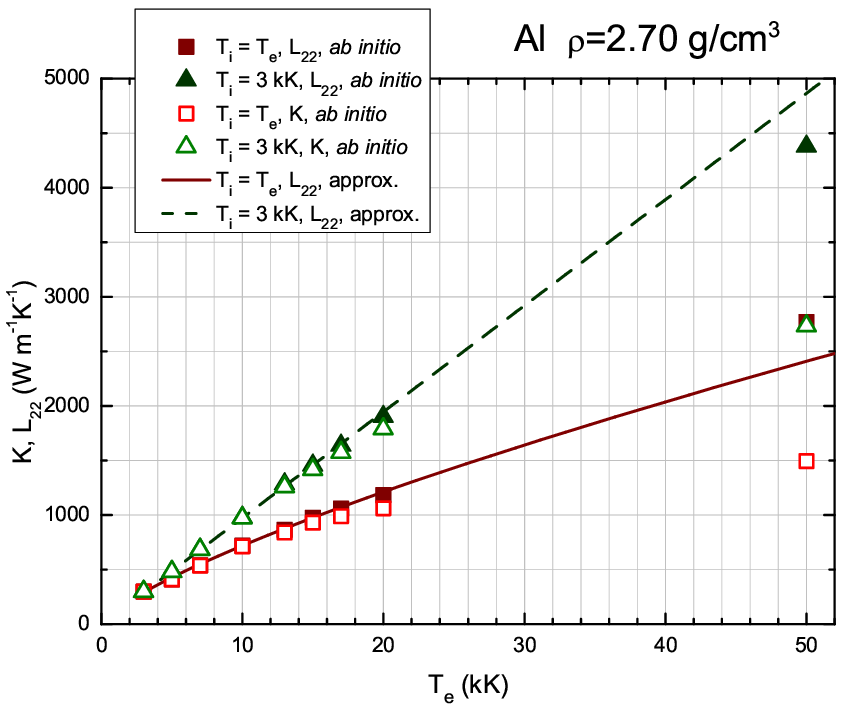}
	b)\includegraphics[width=0.95\columnwidth]{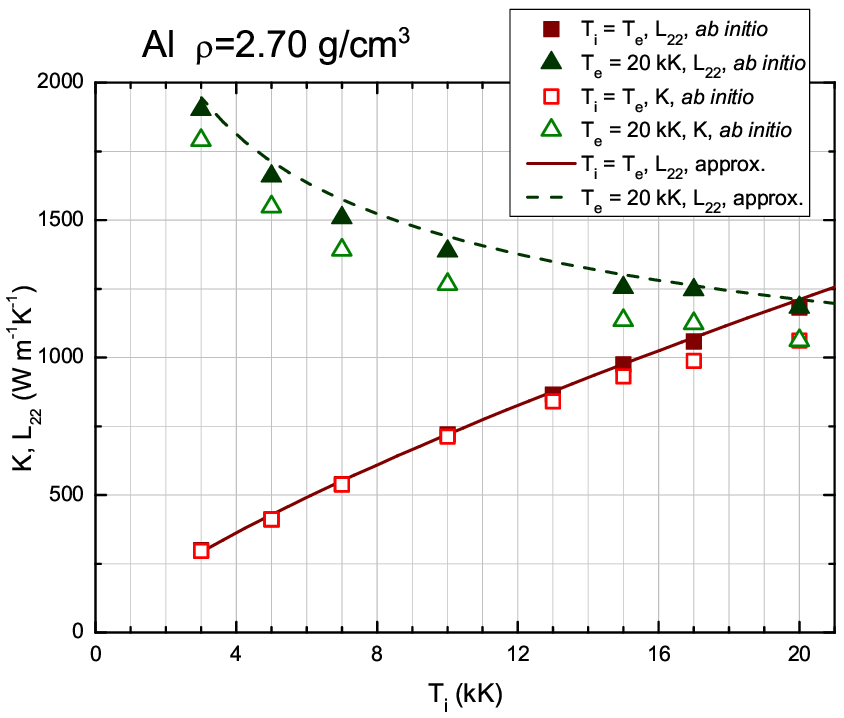}
	\caption{(Color online) Temperature dependences of the thermal conductivity and the $L_{22}$ Onsager coefficient. Filled symbols---$L_{22}$ Onsager coefficient, empty symbols---thermal conductivity. a) Dependences on the electron temperature $T_e$. Squares---equilibrium case $T_i=T_e$; triangles---non-equilibrium case, $T_i=3$~kK, $T_e$ is varied. Solid line---approximation (\ref{Eq:L22_approx}) in equilibrium case, dashed line---approximation (\ref{Eq:L22_approx}) at fixed $T_i=3$~kK. b) Dependences on the ion temperature $T_i$. Squares---equilibrium case $T_e=T_i$; triangles---non-equilibrium case, $T_e=20$~kK, $T_i$ is varied. Solid line---approximation (\ref{Eq:L22_approx}) in equilibrium case, dashed line---approximation (\ref{Eq:L22_approx}) at fixed $T_e=20$~kK.}
	\label{Fig:Al_K_L22}
\end{figure*}

The temperature dependences of the thermal conductivity are shown in Fig.~\ref{Fig:Al_K_L22}.

The dependences of the thermal conductivity on the electron temperature $T_e$ are shown in Fig.~\ref{Fig:Al_K_L22}a. In the equilibrium case, $T_i=T_e$, the thermal conductivity increases as the temperature grows. In the non-equilibrium case $T_i$ is kept fixed at 3~kK, and thermal conductivity increases even more rapidly.

The dependences of the thermal conductivity on the ion temperature $T_i$ are shown in Fig.~\ref{Fig:Al_K_L22}b. In the equilibrium case $T_e=T_i$ we have the same dependence as in Fig.~\ref{Fig:Al_K_L22}a. If $T_e$ is kept fixed at 20~kK and $T_i$ is varied from 3~kK up to 20~kK, thermal conductivity decreases.

The dependences of the thermoelectric term on the electron temperature $T_e$ are shown in Fig.~\ref{Fig:Al_K_L22}a. In the equilibrium case, the contribution of the thermoelectric term grows as the temperature increases. If $T_i$ is kept fixed the thermoelectric contribution also increases as $T_e$ grows.

The dependences of the thermoelectric term on the ion temperature $T_i$ are shown in Fig.~\ref{Fig:Al_K_L22}b. The equilibrium results in Fig.~\ref{Fig:Al_K_L22}b are obviously the same as in Fig.~\ref{Fig:Al_K_L22}a: the thermoelectric term increases with the temperature growth. If $T_e$ is kept fixed the thermoelectric term is almost independent on the ion temperature $T_i$.

\begin{figure*}
  a)\includegraphics[width=0.95\columnwidth]{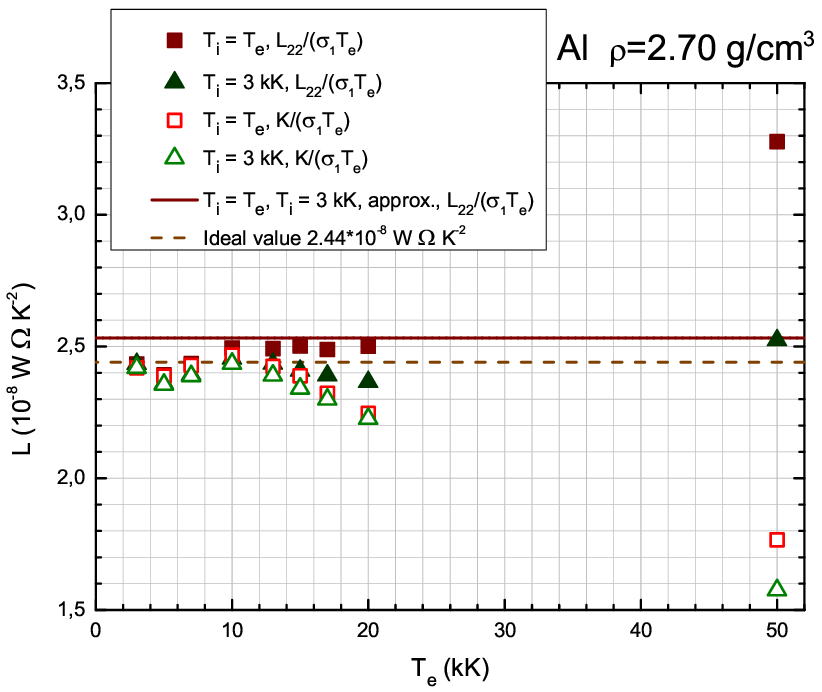}
  b)\includegraphics[width=0.95\columnwidth]{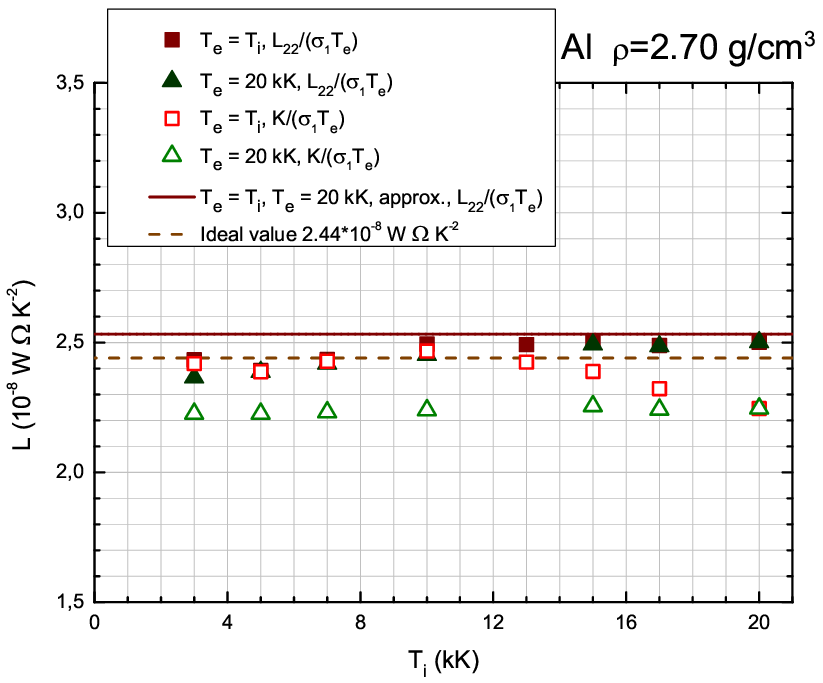}
  \caption{(Color online) Temperature dependences of the calculated Lorenz number. Filled symbols---the Lorenz numbers calculated according to the $\frac{L_{22}}{\sigma_{1_\mathrm{DC}}\cdot T_e}$ expression, empty symbols---the expression $\frac{K}{\sigma_{1_\mathrm{DC}}\cdot T_e}$. a) The dependences on the electron temperature $T_e$. Squares---equilibrium case $T_i=T_e$; triangles---non-equilibrium case, $T_i=3$~kK, $T_e$ is varied. Solid line---approximation of the Lorenz number (\ref{Eq:Lorenz_approx}) $L=2.533\times 10^{-8}~\mathrm{W~\Omega~K^{-2}}$; dashed line---the ideal value of the Lorenz number $L=2.44\times 10^{-8}~\mathrm{W~\Omega~K^{-2}}$. b) Dependences on the ion temperature $T_i$. Squares---equilibrium case $T_e=T_i$; triangles---non-equilibrium case, $T_e=20$~kK, $T_i$ is varied. Solid line---approximation of the Lorenz number (\ref{Eq:Lorenz_approx}); dashed line---the ideal value of the Lorenz number.}
  \label{Fig:Al_Lorenz}
\end{figure*}

If thermal conductivity and static electrical conductivity are known, the Wiedemann-Franz law may be checked. The Wiedemann-Franz law was discovered at rather low temperatures. At those conditions $T_e=T_i$ and $K\approx L_{22}$; the law is presented in form (\ref{Eq:WiedemannFranz}). In our case $T_e$ and $T_i$ may differ; at rather high temperatures $K$ and $L_{22}$ differ also (Fig.~\ref{Fig:Al_K_L22}). Therefore, the question arises, what expression for the Lorenz number should be used. In this work we tested the expressions $\frac{K}{\sigma_{1_\mathrm{DC}}\cdot T_e}$ and $\frac{L_{22}}{\sigma_{1_\mathrm{DC}}\cdot T_e}$. The results for the Lorenz number are shown in Fig.~\ref{Fig:Al_Lorenz}.

The dependences of the Lorenz number on the electron temperature $T_e$ are shown in Fig.~\ref{Fig:Al_Lorenz}a. The Lorenz numbers for temperatures $T_e$ lower than 20~kK are close to the ideal value, both in equilibrium and non-equilibrium cases, both calculated using $L_{22}$ and $K$. The relative difference of all the points from the ideal value is within 9\%. This is smaller than the estimated error calculation of about 20\% (see section~\ref{Sec:Parameters} and paper \cite{Knyazev:COMMAT:2013}). Thus we can conclude that the Wiedemann-Franz law is approximately valid for the points considered. The Lorenz numbers calculated using the $L_{22}$ coefficient seem to be closer to the ideal value (discrepancy not larger than 3\%), than those calculated with $K$ (discrepancy not larger than 9\%).

The dependences of the Lorenz number on the ion temperature $T_i$ are shown in Fig.~\ref{Fig:Al_Lorenz}b. All the points presented are close to the ideal value (the discrepancy less than the estimated error of calculation). The points calculated using $L_{22}$ values are closer to the ideal value (discrepancy not larger than 3\%), than those calculated with $K$ (discrepancy not larger than 9\%).

The results obtained lead us to the following conclusions for liquid aluminum at normal density for the temperatures 3~kK~$\leq T_i \leq T_e \leq 20$~kK. 

The optical properties do not depend on $T_e$, but only on $T_i$. Particularly, in the equilibrium case the static electrical conductivity decreases as the temperature grows. And this decrease is entirely due to the $T_i$ growth. In the non-equilibrium case the static conductivity is fully determined by $T_i$.

The behavior of the thermal conductivity is determined by the behavior of the $L_{22}$ Onsager coefficient. The $L_{22}$ Onsager coefficient depends both on $T_e$ and $T_i$. The growth of $T_e$ leads to the increase of $L_{22}$, growth of $T_i$ leads to the decrease of $L_{22}$. Consequently, in the equilibrium case, we have the struggle of two opposite mechanisms. The influence of $T_e$ turns out to be more powerful and $L_{22}$ still increases as the temperature grows. In the non-equilibrium case with $T_e>T_i$, typical for the femtosecond laser heating, the thermal conductivity is larger than in the equilibrium case with the same $T_i$. The contribution of the thermoelectric term is rather small (the maximum relative contribution is 10\% at $T_i=T_e=20$~kK). The thermoelectric term does not depend on $T_i$ and increases with the growth of $T_e$.

The values of the Lorenz numbers calculated according to the formulas $\frac{K}{\sigma_{1_\mathrm{DC}}\cdot T_e}$ and $\frac{L_{22}}{\sigma_{1_\mathrm{DC}}\cdot T_e}$ are close to the ideal value (discrepancy lower than the calculation error). The values calculated using the $L_{22}$ value seem to be even closer to the ideal value than those calculated using thermal conductivity $K$.

Two points at normal density---$T_i=T_e=50$~kK and $T_i=3$~kK, $T_e=50$~kK---were calculated for the temperatures larger than 20~kK. At these temperatures some of the conclusions drawn above become incorrect. The optical properties and static electrical conductivity start to depend on $T_e$ (Fig.~\ref{Fig:Al_resigma}b and Fig.~\ref{Fig:Al_resigmaDC}a). The thermoelectric term turns out to give significant contribution to the thermal conductivity (Fig.~\ref{Fig:Al_K_L22}a). The Lorenz number becomes significantly different from the ideal value (Fig.~\ref{Fig:Al_Lorenz}a).

\subsection{The construction of approximation}
\label{Subsec:Results:Approximation}

We have also built the approximation based on the results of our \textit{ab initio} calculation. The points with the temperatures 3~kK~$\leq T_i \leq T_e \leq 20$~kK were used for the construction of the approximation. The procedure was as follows. 

First of all, the results on static electrical conductivity and the $L_{22}$ Onsager coefficient (Figs.~(\ref{Fig:Al_resigmaDC})-(\ref{Fig:Al_K_L22})) were plotted in the double logarithmic scale. In this scale all the dependences were succesfully fitted by the straight lines, both slope and intercept were adjusted. Hereafter all linear fits were performed by the "Fit Linear" option of the Origin software. The slope of the line equals therefore the power $\alpha$ in the approximation $\propto T_{e,i}^\alpha$. The following approximations were obtained:
\begin{eqnarray}
\sigma_{1_\mathrm{DC}}(T)|_{T_e=T_i}\propto\frac{1}{T^{0.279}},
\\
L_{22}(T)|_{T_e=T_i}\propto T^{0.74202},
\\
L_{22}(T_e)|_{T_i=3~\mathrm{kK}}\propto T_e^{0.986},
\\
L_{22}(T_i)|_{T_e=20~\mathrm{kK}}\propto\frac{1}{T_i^{0.24718}}.
\end{eqnarray}

The powers obtained are close to the values 0.25, 0.75, 1. This, together with the conclusions of the \textit{ab initio} calculations drawn in the previous subsection, brings to the mind the idea to test the following form of approximation:
\begin{eqnarray}
\sigma_{1_\mathrm{DC}}(T_i,T_e)=\frac{A}{T_i^{0.25}},
\\
L_{22}(T_i,T_e)=B\frac{T_e}{T_i^{0.25}},
\\
L=\frac{L_{22}}{\sigma_{1_\mathrm{DC}}\cdot T_e}=\frac{B}{A}.
\label{Eq:Lorenz_approx_idea}
\end{eqnarray}
Here $A$ and $B$ are adjustable constants to be determined. It is worth noting, this form of approximation yields the correct Wiedemann-Franz law (\ref{Eq:Lorenz_approx_idea}).

The coefficients $A$ and $B$ were determined as follows. The \textit{ab initio} $\sigma_{1_\mathrm{DC}}(T)|_{T_e=T_i}$ dependence was plotted in the double logarithmic scale and fitted by the straight line. Now the slope was set exactly to -0.25 and the intercept only was adjusted. The intercept yielded the $A$ value. Then $L_{22}(T_e)|_{T_i=3~\mathrm{kK}}$ \textit{ab initio} dependence was plotted in double logarithmic scale and fitted by the straight line. The slope was set exactly to 1 and intercept was adjusted. The intercept yielded the $B$ value, taking into account the ionic temperature of 3000~K. 

Thus we come to the following approximation of the transport properties:
\begin{eqnarray} \sigma_{1_\mathrm{DC}}(T_i,T_e)\left[\mathrm{\Omega^{-1}m^{-1}}\right]=\frac{2.844\cdot10^7}{\left(T_i[\mathrm{K}]\right)^{0.25}},
\label{Eq:ResigmaDC_approx}
\\
L_{22}(T_i,T_e)[\mathrm{W~m^{-1}K^{-1}}]=0.720\cdot\frac{T_e[\mathrm{K}]}{\left(T_i[\mathrm[K]]\right)^{0.25}},
\label{Eq:L22_approx}
\\
L=2.533\cdot10^{-8} \left[\mathrm{W~\Omega~K^{-2}}\right].
\label{Eq:Lorenz_approx}
\end{eqnarray}

The comparison of the approximations with the \textit{ab initio} data is presented in Figs.~(\ref{Fig:Al_resigmaDC})-(\ref{Fig:Al_Lorenz}). The comparison is to be performed only for the points with the temperatures 3~kK~$\leq T_i \leq T_e \leq 20$~kK. The error of the static electrical conductivity approximation is not larger than 7\% (Fig.~\ref{Fig:Al_resigmaDC}). The error of the thermal conductivity approximation is not larger than 5\% (Fig.~\ref{Fig:Al_K_L22}). The discrepancy between the Lorenz numbers calculated using the $L_{22}$ coefficient and the approximate value (\ref{Eq:Lorenz_approx}) is not larger than 8\% (Fig.~\ref{Fig:Al_Lorenz}). 

Thus the approximation rather well reproducing the transport properties was constructed. Only thermoelectric coefficients $L_{12}$ and $L_{21}$ and thermoelectric contribution to the thermal conductivity were not described for the following reasons. The constructed approximation works well for the temperatures less than 20~kK. At those temperatures the thermoelectric term is small, the precision of its calculation is low, and it can hardly be investigated properly (it is also not necessary due to its negligible contribution). At the temperatures higher than 20~kK the thermoelectric term is evidently important (Fig.~\ref{Fig:Al_K_L22}). But at those temperatures the current approximation is not valid and should be further reworked.

We have also extended the approximation to describe the optical properties.

The dynamic electrical conductivity was approximated by the Drude formula:
\begin{equation}
\sigma_1(\omega)=\frac{\sigma_{1_\mathrm{DC}}}{1+\omega^2\tau^2}.
\label{Eq:Drude}
\end{equation}
The Drude formula was linearized:
\begin{equation}
\frac{1}{\sigma_1(\omega)}=\frac{1}{\sigma_{1_\mathrm{DC}}}+\frac{\tau^2}{\sigma_{1_\mathrm{DC}}}\omega^2.
\end{equation}

The frequency dependences of the dynamic electrical conductivity for each temperature point 3~kK~$\leq T_i=T_e\leq20$~kK were plotted in $\frac{1}{\sigma_1(\omega)}$ vs. $\omega^2$ plot. The \textit{ab initio} data were fitted by the straight line. The intercept was kept fixed at the values $\frac{1}{\sigma_{1_\mathrm{DC}}}$, obtained previously by the linear extrapolation to zero frequency (see Section~\ref{Sec:Technique}). The slope was adjusted to fit the \textit{ab initio} data. The effective relaxation time $\tau$ was expressed via the slope of the fitting straight line. Thus we get a number of $\tau(T)$ points.

The set of $\tau(T)$ values along with the corresponding $\sigma_{1_\mathrm{DC}}(T)$ values already provides an approximation. It enables the calculation of the dynamic electrical conductivity for the given temperature according to the formula (\ref{Eq:Drude}). The error of such an approximation is not larger than 9\% for all frequencies $\omega\leq10$~eV. However it is more convenient to approximate $\tau(T)$ dependence by the smooth curve as it has been done previously for $\sigma_{1_\mathrm{DC}}(T)$. Having plotted $\tau(T)$ in the double logarithmic scale and fitted it by straight line, we have received the approximation:
\begin{equation}
\tau(T)|_{T_i=T_e}\propto\frac{1}{T^{0.288}}.
\end{equation}

This leads us to the idea to approximate $\tau(T)$ dependence by the expression:
\begin{equation}
\tau(T)|_{T_i=T_e}=\frac{C}{T^{0.25}}.
\end{equation}
This particular choice of the power of 0.25 possesses the following advantage. This power is the same as one in the approximation of the static electrical conductivity (\ref{Eq:ResigmaDC_approx}). So $\tau(T)$ and $\sigma_{1_\mathrm{DC}}(T)$ become proportional to each other. This is natural for the Drude approach \cite{Ashcroft:1976}, where the ratio of $\sigma_{1_\mathrm{DC}}(T)$ and $\tau(T)$ is a constant (for a given density) connected with the plasma frequency:
\begin{equation}
\frac{\sigma_{1_\mathrm{DC}}(T)}{\tau(T)}=\omega^2_{\mathrm{pl}}\varepsilon_0.
\end{equation}
The latter formula is presented in SI units and $\varepsilon_0$ is the dielectric permittivity of vacuum.

One more time the dependence $\tau(T)$ was plotted in the double logarithmic scale. The points were fitted by the straight line. This time the slope was kept fixed at -0.25 and only the intercept was adjucted. This yielded us the $C$ value. Thus we receive the approximation for the relaxation time $\tau$:
\begin{equation}
\tau\left[\mathrm{eV^{-1}}\right]=\frac{7.876}{\left(T_i[\mathrm{K}]\right)^{0.25}},
\label{Eq:tau_approx}
\end{equation}
and the final approximation for the dynamic electrical conductivity:
\begin{equation}
\sigma_1(T_i,T_e,\omega)\left[\mathrm{\Omega^{-1}m^{-1}}\right]=\frac{\frac{2.844\cdot10^7}{\left(T_i[\mathrm{K}]\right)^{0.25}}}{1+\left(\frac{7.876}{\left(T_i[\mathrm{K}]\right)^{0.25}}\right)^2\left(\omega\left[\mathrm{eV}\right]\right)^2}.
\label{Eq:Resigma_approx}
\end{equation}
The approximation also gives the value of plasma frequency:
\begin{equation}
\omega_{\mathrm{pl}}=16.83~\text{eV}.
\label{Eq:omegapl_approx}
\end{equation}

The discrepancy between the approximation and the results of the \textit{ab initio} calculation is not larger than 13\% for all points at the normal density, 3~kK~$\leq T_i \leq T_e\leq20$~kK, $\omega\leq10$~eV. The latter restrictions determine the region where the approximations (\ref{Eq:ResigmaDC_approx})--(\ref{Eq:Lorenz_approx}), 
(\ref{Eq:tau_approx})--(\ref{Eq:omegapl_approx}) are valid. 

\begin{figure}
	\includegraphics[width=0.95\columnwidth]{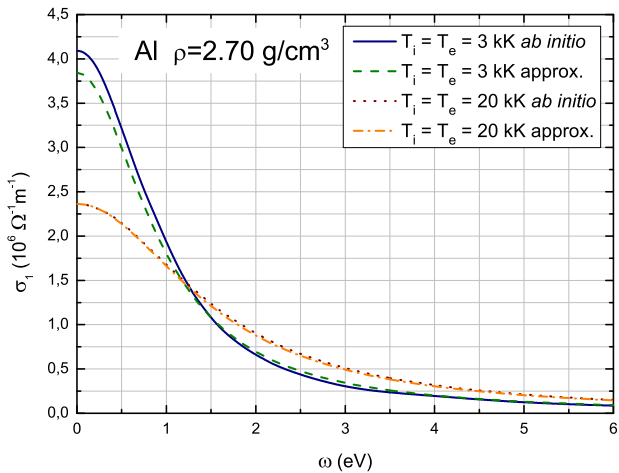}
	\caption{(Color online) Comparison of the \textit{ab initio} data on the optical properties with the approximation (\ref{Eq:Resigma_approx}). Normal density, equilibrium case $T_i=T_e$. Solid line---\textit{ab initio} data, 3~kK; dotted line---\textit{ab initio} data, 20~kK. Dashed line---approximation (\ref{Eq:Resigma_approx}), 3~kK; dash-dot line---approximation (\ref{Eq:Resigma_approx}), 20~kK.}
	\label{Fig:Al_resigma_approx}
\end{figure}

The example of the comparison of the \textit{ab initio} data with the approximation (\ref{Eq:Resigma_approx}) is presented in Fig.~\ref{Fig:Al_resigma_approx}. When the temperature is changed from 3~kK to 20~kK, at low frequencies the dynamic electrical conductivity increases almost by a factor of 2, at high ones---decreases by a factor of 2. The difference of the approximation from the \textit{ab initio} curves is not higher than 13\% for all the temperatures and frequencies considered. So the error of approximation is significantly lower than the characteristic changes of the curves under temperature variation. This is an argument for the applicability of the approximation developed. The error of approximation is lower than the estimated error of our \textit{ab initio} calculation (see Section~\ref{Sec:Parameters} and paper \cite{Knyazev:COMMAT:2013}), this also supports the approximation constructed.

\subsection{Comparison with other models}
\label{Subsec:Results:Models}

\subsubsection{The Drude model}

The simplest model, frequently used for the description of two-temperature electrical and thermal conductivity is based on the Drude approach. The thermal conductivity is expressed via the volume-specific heat capacity $C_e$ and the electron relaxation time $\tau_e$ \cite{Ashcroft:1976, Ivanov:PRB:2003}:
\begin{equation}
K=\frac{1}{3}v^2_FC_e\tau_e,
\label{Eq:K_Drude}
\end{equation}
where $v_F$ is the Fermi velocity. The static electrical conductivity is expressed via the electron relaxation time $\tau_e$ and electron concentration $n_e$ \cite{Ashcroft:1976}:
\begin{equation}
\sigma_{1_\mathrm{DC}}=\frac{n_e e^2\tau_e}{m_e},
\label{Eq:ResigmaDC_Drude}
\end{equation}
where $m_e$ and $e$ are the electron mass and charge, respectively. 

The temperature dependence of the heat capacity is considered to be linear \cite{Ivanov:PRB:2003}, $C_e=\gamma T_e$, that corresponds to the ideal Fermi-gas at $T_e\ll T_F$, $T_F$ is the Fermi temperature. Simplest assumptions are made for the electron relaxation time $\tau_e$ \cite{Ivanov:PRB:2003}: $1/\tau_e=1/\tau_{e-e}+1/\tau_{e-ph}$, where $\tau_{e-e}$ and $\tau_{e-ph}$ are the electron-electron and electron-phonon relaxation times, respectively; their 
temperature dependencies are $1/\tau_{e-ph}=AT_i$ and $1/\tau_{e-e}=BT_e^2$ expressions are used to describe the temperature dependences of the relaxation times. The expression for $K$ is obtained \cite{Ivanov:PRB:2003}:
\begin{equation}
K=\frac{1}{3}v^2_F\frac{\gamma T_e}{AT_i+BT_e^2}.
\label{Eq:K_Ivanov}
\end{equation}
Here $A$ and $B$ are dimensional coefficients. The assumption $1/\tau_{e-ph}=AT_i$ is based on the linear dependence of the number of phonons on the ion temperature. The very concept of phonons is based on the idea of a solid phase and small displacements of atoms from their equilibrium positions. Expression (\ref{Eq:K_Ivanov}) is used in the paper \cite{Ivanov:PRB:2003} only for the crystalline phase and the temperatures slightly above melting. Also, paper \cite{Ivanov:PRB:2003} claims that the term $1/\tau_{e-ph}=BT_e^2$ may be neglected for the temperatures significantly below the Fermi temperature.

We may obtain the similar expression for the static electrical conductivity:
\begin{equation}
\sigma_{1_\mathrm{DC}}=\frac{n_e e^2}{m_e}\frac{1}{AT_i+CT_e^2}.
\label{Eq:Resigma_Ivanov}
\end{equation}
The electron-electron collisions in the latter expression are described by the term $CT_e^2$. The coefficient $C$ is smaller or equal than the coefficient $B$ in expression (\ref{Eq:K_Ivanov}). This difference is introduced to take into account the probably different influences of the electron-electron collisions on thermal conductivity and electrical conductivity \cite{Inogamov:JETP:2010}.

Expressions (\ref{Eq:K_Ivanov}) and (\ref{Eq:Resigma_Ivanov}), with electron-electron collision term neglected, ensure the validity of the Wiedemann-Franz law in the form:
\begin{equation}
\frac{K}{\sigma_{1_\mathrm{DC}}\cdot T_e} = \text{const}.
\end{equation}

In this work we have performed \textit{ab initio} calculations of static electrical conductivity and thermal conductivity and constructed the approximations based on the calculations. Therefore we may verify expressions (\ref{Eq:K_Ivanov})--(\ref{Eq:Resigma_Ivanov}). We work at $T_i$ significantly above melting, and $T_e$ significantly below the Fermi temperature $T_F$ (for aluminum at normal density it is about 135~kK, taking 3 electrons into account).

The statement on the negligibility of the $CT_e^2$ in (\ref{Eq:Resigma_Ivanov}) is proved by our data: the static electrical conductivity does not depend on $T_e$. In that case according to (\ref{Eq:Resigma_Ivanov})
$\sigma_{1_\mathrm{DC}}$ should be $\propto T_i^{-1}$. This is not proved by our data: $\sigma_{1_\mathrm{DC}}$ decreases slower, as $\propto T_i^{-0.25}$. We do not assign any physical meaning to our approximation (\ref{Eq:Resigma_approx}), the coefficients and the powers of the approximation may be slightly different, but the lower rate of the $\sigma_{1_\mathrm{DC}}$ decrease is obvious. We have also tried to approximate our \textit{ab initio} data by the expression $\sigma_{1_\mathrm{DC}}\propto\frac{1}{AT_i+\mathrm{const}}$ with no result. Thus the expression (\ref{Eq:Resigma_Ivanov}) is not supported by our data. 

The same conclusions may be drawn for the thermal conductivity, if the ratio $K/T_e$ is investigated and  expression (\ref{Eq:K_Ivanov}) is taken as a model for $K$.

Thus we have checked, that our \textit{ab initio} data can not be described by the Drude model with these particular expressions for the relaxation times (\ref{Eq:K_Ivanov})--(\ref{Eq:Resigma_Ivanov}). However our data do not contradict to the Drude model (\ref{Eq:K_Drude})--(\ref{Eq:ResigmaDC_Drude}) with  expression (\ref{Eq:tau_approx}) for the relaxation time. The electron concentration $n_e$ in (\ref{Eq:ResigmaDC_Drude}) is connected with the atom concentration $n_{a}$:
\begin{equation}
n_e=Zn_{a}=Z\frac{\rho}{\mu}N_{A}.
\end{equation}
Here $\rho$ is the density of material (normal density of aluminum is 2.70~g/cm$^3$), $\mu$ is its molar mass (26.98~g/mol for aluminum), $N_A$ is the Avogadro constant. The ion charge $Z$ is taken to be 3. The volume-specific heat capacity at constant volume $C_e$ in (\ref{Eq:K_Drude}) is calculated via the low-temperature asymptotic formula for the degenerate ideal Fermi-gas \cite{LandauLifshitz:V}:
\begin{equation}
C_e=\left(\frac{\pi}{3}\right)^{2/3}\frac{m_e}{\hbar^2}k_B^2 n_e^{1/3} T_e.
\label{Eq:ce_degenerate}
\end{equation}
Here $k_B$ is the Boltzmann constant, $\hbar$ is the reduced Planck constant. The square of the Fermi velocity $v_F^2$ in (\ref{Eq:K_Drude}) is calculated for the ideal Fermi gas at zero temperature:
\begin{equation}
v_F^2=\frac{E_F}{2m_e}=\frac{\hbar^2}{4m_e^2}(3\pi^2)^{2/3}n_e^{2/3}.
\label{Eq:velocity_squared}
\end{equation}
\begin{figure}
a)~\includegraphics[width=0.95\columnwidth]{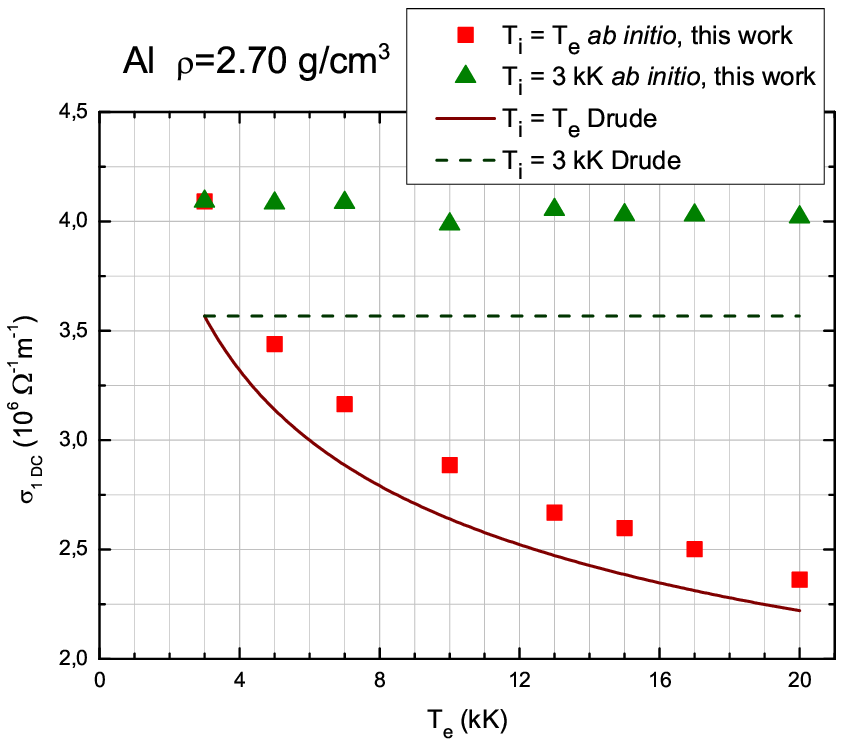}
b)~\includegraphics[width=0.95\columnwidth]{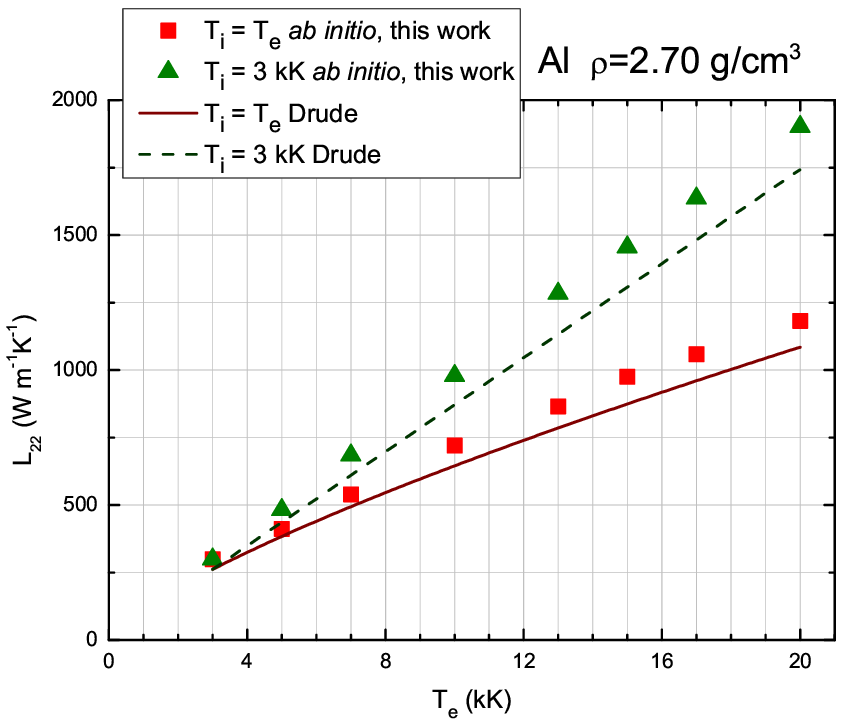}
\caption{(Color online) The comparison of the \textit{ab initio} data with the results of the Drude model, $Z=3$. Squares---\textit{ab initio} data, equilibrium case, $T_i=T_e$; triangles---\textit{ab initio} data, non-equilibrium case, $T_i=3$~kK. Solid line---the Drude model, equilibrium case, $T_i=T_e$; dashed line---non-equilibrium case, $T_i=3$~kK. a) Static electrical conductivity. b) The $L_{22}$ Onsager coefficient.}
\label{Fig:Drude_model}
\end{figure}
The results of the comparison of the \textit{ab initio} data with the Drude model described above are shown in Fig.~\ref{Fig:Drude_model}. The comparison is performed only for 3~kK~$\leq T_i\leq T_e \leq20$~kK, because only in this temperature range the approximation for the relaxation time (\ref{Eq:tau_approx}) is valid; $T_e$ should be significantly below the Fermi temperature to use expressions (\ref{Eq:ce_degenerate})-(\ref{Eq:velocity_squared}).

The results on the static electrical conductivity are shown in Fig.~\ref{Fig:Drude_model}a. The qualitative agreement of the Drude model with the \textit{ab initio} data is present. The results of the Drude model are sistematically low than the \textit{ab initio} data. 

Using the approximate value of the plasma frequency (\ref{Eq:omegapl_approx}) the effective charge $Z_{\mathrm{eff}}$ may be calculated by the formula:
\begin{equation}
Z_{\mathrm{eff}}=\frac{\omega_{\mathrm{pl~approx}}^2\varepsilon_0}{\frac{n_{a}e^2}{m_e}}.
\end{equation}
The calculation yields $Z_{\mathrm{eff}}=3.23$. If the Drude model is calculated with this effective charge, we immediately receive the approximation (\ref{Eq:ResigmaDC_approx}) for static electrical conductivity and better agreement with the \textit{ab initio} data (compare Fig.~\ref{Fig:Al_resigmaDC}a and Fig.~\ref{Fig:Drude_model}a). It is also worth noting, that $Z_{\mathrm{eff}}=3.23$ value is not that far from 3.

The results on the $L_{22}$ Onsager coefficient are shown in Fig.~\ref{Fig:Drude_model}b. The derivation of the Drude expression (\ref{Eq:K_Drude}) presented in \cite{Ashcroft:1976} neglects the thermoelectric terms. Consequently it is more correct to use $L_{22}$ \textit{ab initio} values rather than $K$ for the comparison with the Drude model.

Thus the results of the \textit{ab initio} calculation may be satisfactorily described by the Drude model, but the frequency of the electron-ion collisions should not grow $\propto T_i$. The rate of its growth should be smaller, in this work it was approximated as $\propto T_i^{0.25}$.

\subsubsection{The Anisimov model}

Another analytical model, describing the two-temperature thermal conductivity was proposed in paper \cite{Anisimov:SPIE:1997}:
\begin{equation}
K=C\cdot\frac{(t_e^2+0.16)^{5/4}(t_e^2+0.44)}{(t_e^2+0.092)^{1/2}}\frac{t_e}{\beta t_i+t_e^2},
\label{Eq:Anisimov}
\end{equation}
where $t_e=T_e/T_F$ and $t_i=T_i/T_F$ are the dimensionless temperatures of electrons and ions, respectively. $C$ and $\beta$ coefficients may be derived from experimental data. The density dependence is accounted by the $T_F(\rho)$. Hereafter the model (\ref{Eq:Anisimov}) will be called the \textit{Anisimov model}. This model gives the correct high-temperature Spitzer asymptotics $K\propto T_e^{5/2}$ for $T_e\gg T_F$. In this work, however, we will consider the Anisimov model only at low temperatures $T_i\leq T_e \ll T_F$.

The coefficients $C=770$~W~m$^{-1}$K$^{-1}$ and $\beta=1.2$ are presented in paper \cite{Anisimov:JETP:2006}. The coefficient $C$ was chosen to obtain the correct experimental value of thermal conductivity at $T_i=T_e=1$~kK. In our previous work \cite{Knyazev:COMMAT:2013} we achieved good agreement with the experiment at $\rho=2.35$~g/cm$^{-3}$, $T_i=T_e=1$~kK. So in these conditions the results of the Anisimov model are close to ours.

\begin{figure}
a)~\includegraphics[width=0.95\columnwidth]{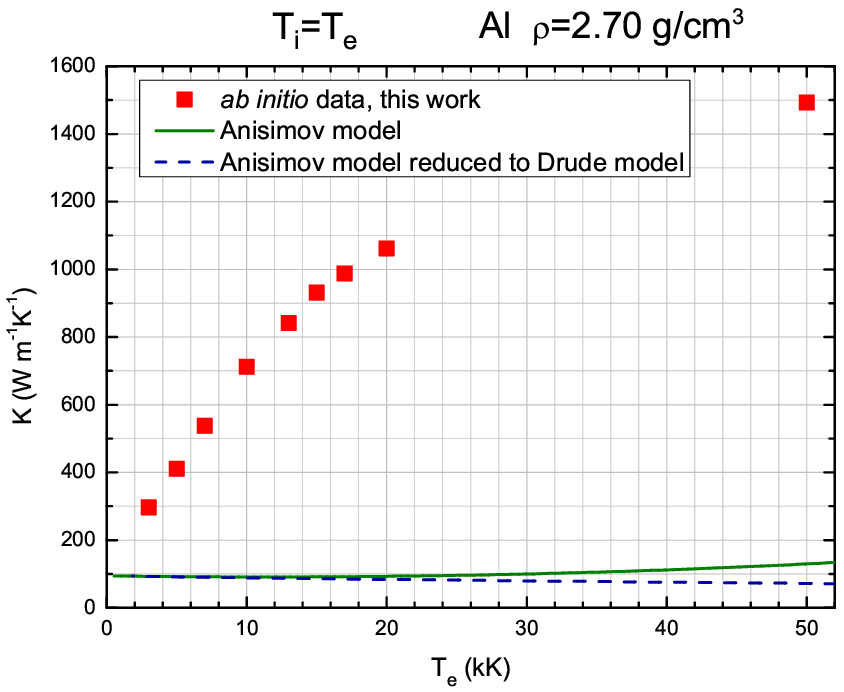}
b)~\includegraphics[width=0.95\columnwidth]{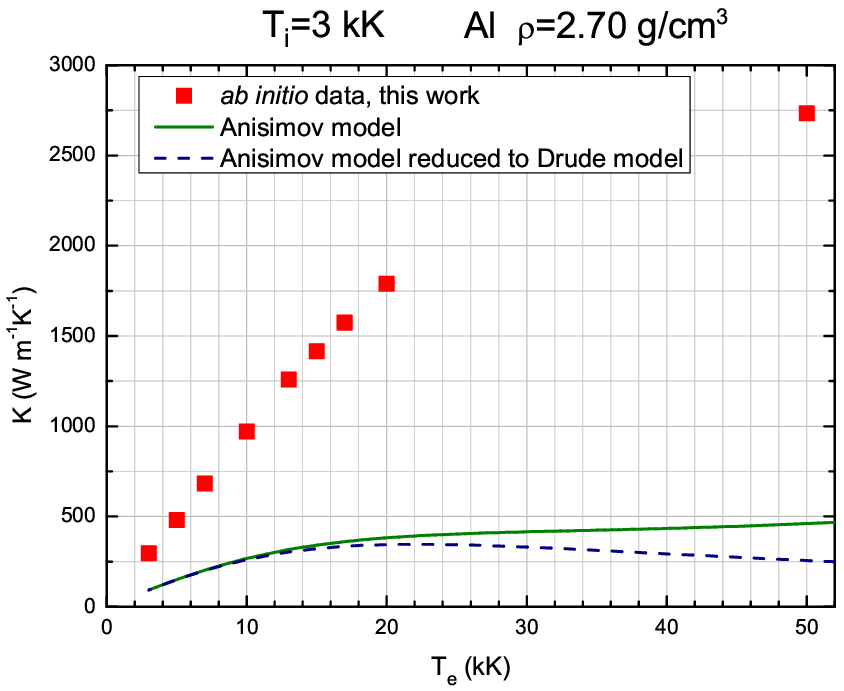}
\caption{(Color online) Comparison of the \textit{ab initio} data on the thermal conductivity with the Anisimov model. Squares---\textit{ab initio} data. Solid line---the Anisimov model (\ref{Eq:Anisimov}), dashed line---the reduced Anisimov model (\ref{Eq:Anisimov_reduced}). a) Equilibrium case $T_i=T_e$. b) Non-equilibrium case $T_i=3$~kK.}
\label{Fig:Anisimov_Inogamov}
\end{figure}

In Fig.~\ref{Fig:Anisimov_Inogamov} we present the comparison of our results with the Anisimov model in equilibrium (Fig.~\ref{Fig:Anisimov_Inogamov}a) and non-equilibrium (Fig.~\ref{Fig:Anisimov_Inogamov}b) cases. It should be mentioned that at $T_e<20$~kK the first factor in (\ref{Eq:Anisimov}) is almost constant, and all the changes in the thermal conductivity may be described by the reduced expression:
\begin{equation}
K=C\cdot\frac{0.16^{5/4}\cdot0.44}{0.092^{1/2}}\frac{t_e}{\beta t_i+t_e^2}.
\label{Eq:Anisimov_reduced}
\end{equation}
This reduced expression is simply the Drude model in the form (\ref{Eq:K_Ivanov}). As before the term $\beta t_i$ in the denominator is due to the electron-phonon collisions, the term $t_e^2$---due to the electron-electron ones. The results of the calculation both according to the full (\ref{Eq:Anisimov}) and reduced (\ref{Eq:Anisimov_reduced}) expressions are shown in Fig.~\ref{Fig:Anisimov_Inogamov}.

In the equilibrium case (Fig.~\ref{Fig:Anisimov_Inogamov}a) the reduced expression yields the values almost equal to that at $T_i=T_e=1$~kK. The electron-phonon collisions significantly dominate electron-electron collisions in (\ref{Eq:Anisimov_reduced}) (their contributions become
equal only at $T_i=T_e=162$~kK). Thus we have $\propto T_e/T_i$ dependence that in the equilibrium case yields the constant thermal conductivity. The constant temperature behavior of the thermal conductivity is a characteristic of the crystalline phase \cite{Ziman:1960}. The Anisimov model shows this behavior even at very high temperatures. Our calculations give the different, increasing dependence. Particularly, at $T_i=T_e=3$~kK our values significantly exceed the results of the Anisimov model.

In the non-equilibrium case our dependences also differ significantly. Already at $T_i=T_e=3$~kK our results are significantly higher due to the reasons mentioned above. At constant $T_i=3$~kK at first the Anisimov model gives the curve that increases $\propto T_e$. Our dependence also has $\propto T_e$ shape, though the slope is different. But already at 22~kK the contributions of the electron-phonon and electron-electron collisions in (\ref{Eq:Anisimov_reduced}) become equal. The term $t_e^2$ in the denominator prevents the Anisimov curve from further increase.

Thus the differences between our results and the Anisimov model are due to the following two reasons. Firstly, the Anisimov model at low temperatures has the dependence $\propto T_e/T_i$, whereas we predict different behavior, $\propto T_e/T_i^{0.25}$. Secondly, in the non-equilibrium case in the Anisimov model the electron-electron collisions become significant already at rather moderate $T_e$ (of about 20~kK).

\subsubsection{The Inogamov-Petrov model}

We have also compared our results with the Inogamov-Petrov model \cite{Inogamov:JETP:2010}.
This approach yields the wide-range expression $K_{sum}^{WR}$ for the thermal conductivity, valid for $T_i,T_e\lesssim330$~kK. It is an interpolation between the condensed matter expression $K_{sum}^c$ and the plasma one $K_{sum}^{pl}$.

The plasma expression is obtained from the Spitzer theory. The contribution of $K_{sum}^{pl}$ to $K_{sum}^{WR}$ is negligible for the temperature region $T_i\leq T_e \leq 50$~kK considered in our work. Thus, we assume $K_{sum}^{WR}\approx K_{sum}^{c}$.

The condensed matter expression $K_{sum}^c$ is composed of the electron-electron contribution $K_{ee}^c$ and the electron-ion one $K_{ei}^c$ according to the formula:
\begin{equation}
	\frac{1}{K_{sum}^c}=\frac{1}{K_{ee}^c}+\frac{1}{K_{ei}^c}.
\end{equation}

The electron-electron contribution $K_{ee}^c$ in the Inogamov-Petrov paper is calculated by solving of the linearized Boltzmann equation, with the $\tau$-approximation for the collision term. To compute the energy dependent electron-electron relaxation time $\tau_{ee}(\epsilon)$ the scattering of all free electrons on each other is considered. The electrons are supposed to interact via the screened Coulomb potential, with the Thomas-Fermi length as the screening radius. In the most of similar works the strongly degenerate case only is considered, and only the collisions of the electrons close to the Fermi sphere are taken into account. In the Inogamov-Petrov work the electrons, far from the Fermi sphere are also taken into account. After $K_{ee}^c$ is obtained, the integral electron-electron relaxation time $\tau_{ee}^c$ may be calculated according to the Drude expression (\ref{Eq:K_Drude}) with the ideal gas expressions (\ref{Eq:ce_degenerate}) and (\ref{Eq:velocity_squared}) for $C_e$ and $v_F^2$, respectively. In the Inogamov-Petrov work the electron-electron relaxation time $\tau_{ee}^c(T_e)$ depends on $T_e$ only. At low temperatures $\tau_{ee}^c(T_e)$ has the common $\sim T_e^{-2}$ asymptotics \cite{Ivanov:PRB:2003}. At high enough temperatures $T_e\gtrsim10$~kK it decreases slower than $\sim T_e^{-2}$ with the temperature growth. The calculation of the $K_{ee}^c$ term is the main contribution of the Inogamov-Petrov work.

To describe the electron-ion term $K_{ei}^c$ Inogamov and Petrov involve the results of the \textit{ab initio} calculation \cite{Recoules:PRB:2005}. \textit{Ab initio} data are approximated by the Drude expression (\ref{Eq:K_Drude}) with ideal gas expressions (\ref{Eq:ce_degenerate}) and (\ref{Eq:velocity_squared}) for $C_e$ and $v_F^2$, respectively. The semiempirical formula
\begin{equation}
	\tau_{ei}^c(T_i)=\frac{A+BT_i-\frac{C}{T_i}}{DT_i}
	\label{Eq:tau_InogamovPetrov}
\end{equation}
is used to approximate the electron-ion relaxation time $\tau_{ei}^c(T_i)$ which depends only on $T_i$. Here $A$, $B$, $C$ and $D$ are dimensional coefficients. The expression different from the common \cite{Ivanov:PRB:2003} expression $\sim T_i^{-1}$ is used: the $BT_i$ term in the numerator of (\ref{Eq:tau_InogamovPetrov}) prevents $\tau_{ei}^c(T_i)$ from the too fast $\sim T_i^{-1}$ decrease. \textit{Ab initio} calculations \cite{Recoules:PRB:2005} provide results for aluminum at $\rho=2.35$~g/cm$^3$ and 1~kK~$\leq T_i=T_e \leq10$~kK.

\begin{figure}
a)~\includegraphics[width=0.95\columnwidth]{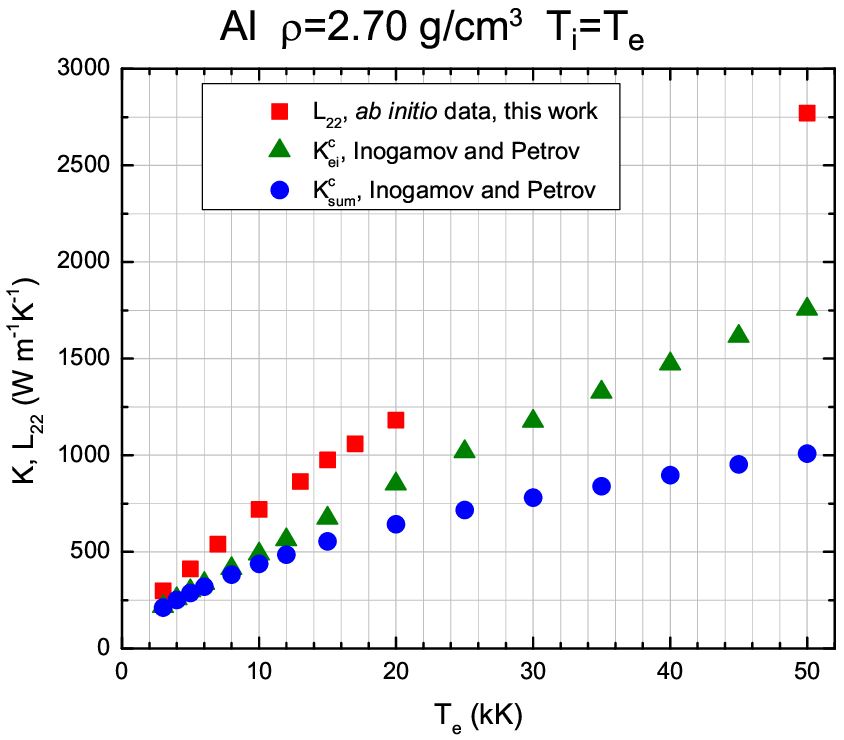}
b)~\includegraphics[width=0.95\columnwidth]{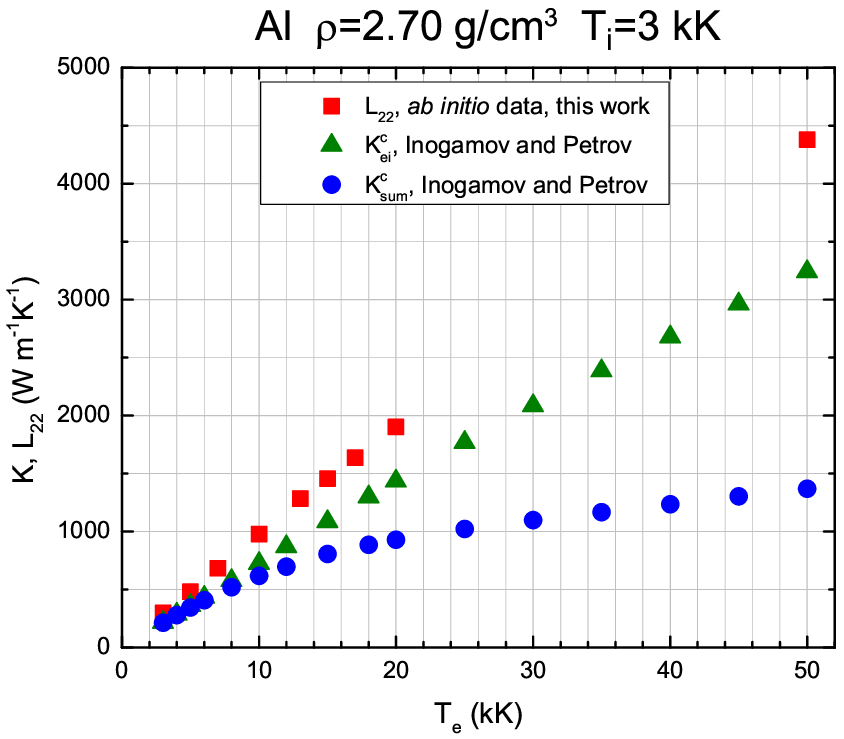}
\caption{(Color online) The comparison of the results of this work with the data of Inogamov and Petrov \cite{Inogamov:JETP:2010}. Squares---\textit{ab initio} calculations of this work. Triangles---\cite{Inogamov:JETP:2010}, $K_{ei}^c$, electron-ion collisions only are taken into account. Circles---\cite{Inogamov:JETP:2010}, $K_{sum}^c$, both electron-ion and electron-electron collisions are taken into account. Liquid aluminum, normal density $\rho=2.70$~g/cm$^3$. a) Equilibrium case $T_i=T_e$. b) Non-equilibrium case $T_i=3$~kK.}
\label {Fig:InogamovPetrov}
\end{figure}

The comparison of our results with the Inogamov-Petrov model is presented in Fig.~\ref{Fig:InogamovPetrov}. Since Inogamov and Petrov do not take the thermoelectric term into account, our results on the $L_{22}$ Onsager coefficient are presented. The comparison will be performed for 3~kK~$\leq T_i \leq T_e \leq20$~kK.

Fig.~\ref{Fig:InogamovPetrov}a shows the results in the equilibrum case $T_i=T_e$.

Since the $K_{ei}^c$ data in the Inogamov-Petrov model is based on the \textit{ab initio} calculation, it should, in principle, coincide with our results. However the discrepancy is present. Both our and $K_{ei}^c$ data are well approximated by the $\sim T^{0.75}$ dependence; the difference is by a constant factor of approximately 1.4. The sources of this discrepancy are easily found. Our results for the density 2.35~g/cm$^3$ are larger than the corresponding values of Recoules \cite{Recoules:PRB:2005} by a factor of approximately 1.15. This discrepancy is most probably due to the used technical parameters (discussed in our previous work \cite{Knyazev:COMMAT:2013}). The approach of Inogamov and Petrov also neglects the difference connected with the change of density from 2.35~g/cm$^3$ to 2.70~g/cm$^3$. This change of the density increases thermal conductivity approximately by a factor 1.2 \cite{Knyazev:COMMAT:2013}. The product of 1.15 and 1.2 yields the resulting factor 1.4 of the discrepancy between our results and $K_{ei}^c$ data of Inogamov and Petrov. However, this is not crucial, because the $K_{ei}^c$ data may be adjusted to reproduce our results.

The contribution of electron-electron collisions may be estimated if we compare $K_{ei}^c$ data with $K_{sum}^c$ (Fig.~\ref{Fig:InogamovPetrov}a). At $T_i=T_e=10$~kK $K_{sum}^c$ is 10\% lower than $K_{ei}^c$; at $T_i=T_e=20$~kK---24\% lower.

The results in the non-equilibrium case $T_i=3$~kK are similar (Fig.~\ref{Fig:InogamovPetrov}b).

Both our results and $K_{ei}^c$ data are well approximated by the $\sim T_e$ dependence; the difference by a constant factor of approximately 1.35 is present (Fig.~\ref{Fig:InogamovPetrov}b). The reasons of this difference are the same as mentioned above for the equilibrium case.

At $T_e=10$~kK $K_{sum}^c$ is 15\% lower than $K_{ei}^c$; at $T_e=20$~kK---35\% lower (Fig.~\ref{Fig:InogamovPetrov}b). If the ion temperature is kept fixed, the influence of the electron-electron collisions is even larger than in the equilibrium case. This may be easily explained. The total relaxation time is determined according to the inverse summation law $1/\tau=1/\tau_{ei}+1/\tau_{ee}$. According to this law the total relaxation time is mainly determined by the smallest of the relaxation times.  If $T_i$ is kept fixed and $T_e$ is increased, $\tau_{ei}$ remains constant, but $\tau_{ee}$ decreases and soon starts to dominate in the total thermal conductivity. In the equilibrium case both $\tau_{ei}$ and $\tau_{ee}$ decrease simultaneously, and the role of $\tau_{ee}$ in the total thermal conductivity is less.

The comparison with the Inogamov-Petrov model brings us to following conclusions. Similar to our work, the Inogamov-Petrov model has the expression for $\tau_{ei}$, that decreases with the temperature growth slower than $\sim T_i^{-1}$. Though the expression of \cite{Inogamov:JETP:2010} differs from ours, it gives the correct temperature dependences of the thermal conductivity (neglecting thermoelectric term). Some quantative discrepancies of $K_{ei}^c$ with our work may be removed by the adjustment of the Inogamov-Petrov model. The contribution of the electron-electron collisions, calculated in \cite{Inogamov:JETP:2010}, turns out to be significant, especially in the non-equilibrium case.

\subsubsection{The Lee-More model}

We have also compared our results with the well-known model of Lee and More \cite{LeeMore:PhysFluids:1984}. 

It is based on the kinetic Boltzmann equation for the free electrons. The interaction of electrons with ions is described by the collision term in the approximation of the relaxation time $\tau_c$ which is defined differently for plasma and liquid. The electron-electron collisions are neglected.

Two different approaches for plasma exist.

\begin{figure}
a)~\includegraphics[width=0.95\columnwidth]{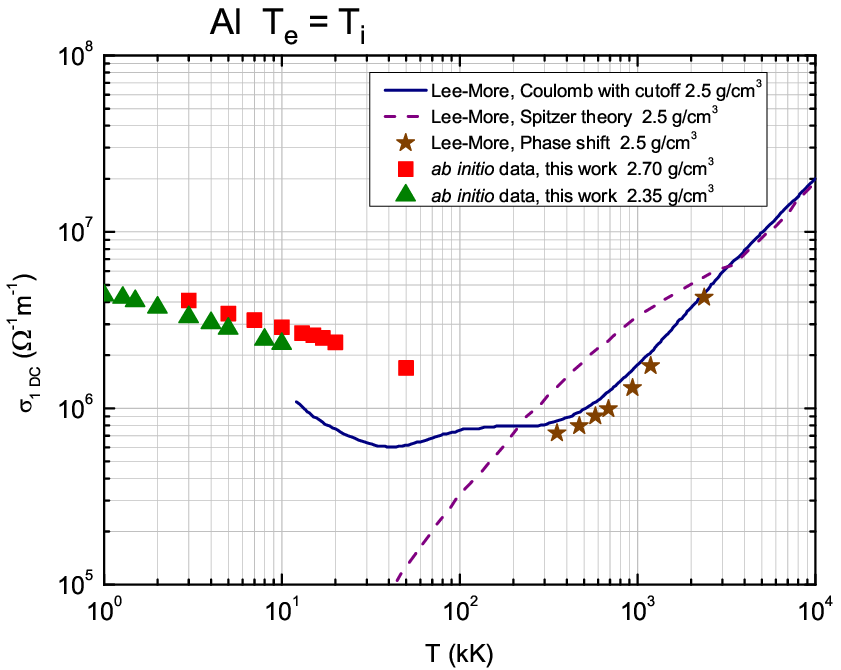}
b)~\includegraphics[width=0.95\columnwidth]{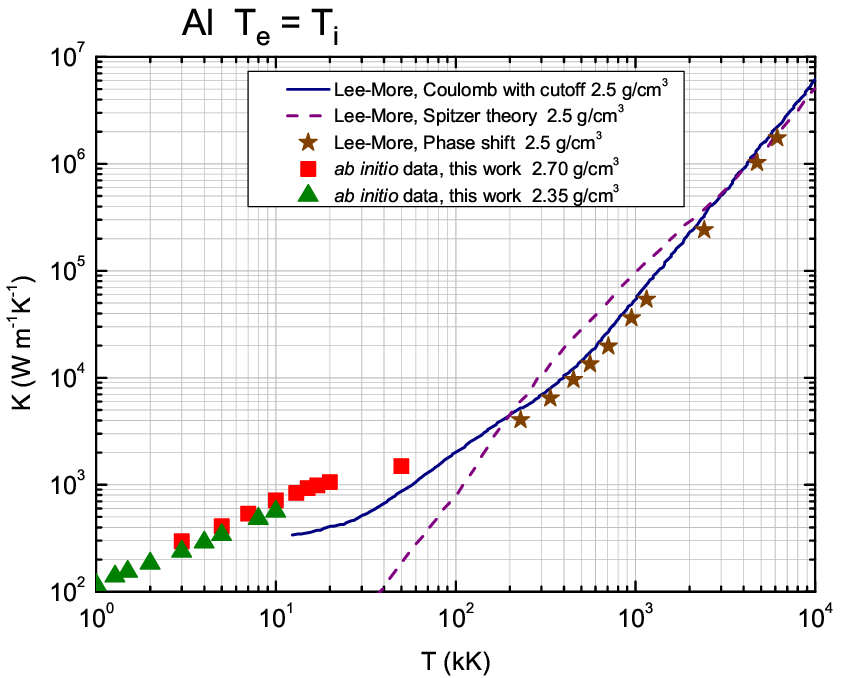}
\caption{(Color online) The comparison of the results of this work with the Lee-More model. Solid line---the Lee-More calculation, based on the Coulomb potential with cutoffs. Dashed line---the Spitzer model as presented in the paper of Lee and More. Stars---the Lee-More results, based on the phase shift calculation for the Thomas-Fermi potential. The density at all Lee-More curves is 2.5~g/cm$^3$. Squares---\textit{ab initio} results of this work, 2.70~g/cm$^3$. Triangles---\textit{ab initio} results of this work, 2.35~g/cm$^3$. a) Static electrical conductivity. b) Thermal conductivity.}
\label{Fig:LeeMore}
\end{figure}

In the first plasma approach the Thomas-Fermi potential of ions is calculated. Then the phase shifts and $\tau_c(\epsilon)$ are obtained from the numerical solution of the Schr\"odinger equation with this potential. The results obtained by this method are depicted by separate points in Fig.~\ref{Fig:LeeMore} and marked with the label "phase shift". The usage of the Thomas-Fermi theory restricts the field of this method by rather high temperatures only.

In the second plasma approach the ion potential is considered to be the Coulomb one with upper and lower cutoffs which depend on the temperatures of electrons and ions and their concentrations. The degeneracy and strong coupling effects are taken into account by choosing appropriate cutoffs. In the second plasma approach the electrical and thermal conductivities are expressed by the integrals over analytic functions. The results obtained by this method are depicted by the solid line in Fig.~\ref{Fig:LeeMore} and marked with the label "Coulomb with cutoffs".

In the liquid case the electron gas is degenerate, the model is reduced to the simple Drude expressions (\ref{Eq:K_Drude})-(\ref{Eq:ResigmaDC_Drude}) with (\ref{Eq:ce_degenerate})-(\ref{Eq:velocity_squared}) expressions for the heat capacity and the square of the Fermi velocity. Also unlike the plasma case in the liquid case the ionic structure factor $S(q)$ is different from unity, therefore the coherent scattering of electrons on ions should be taken into account. Lee and More do not go deep into the investigation of this matter and adopt the Bloch-Gr\"uneisen expression $\tau_c\propto T_i^{-1}$. Moreover, some approximate theories, based on the Lindemann melting criterion, are involved to calculate the coefficient in the Bloch-Gr\"uneisen expression.

The liquid and plasma approaches are conjugated in the intermediate region.

The comparison of our \textit{ab initio} results with the Lee-More model is presented in Fig.~\ref{Fig:LeeMore}. Though the Lee-More model allows different temperatures of electrons and ions, only equilibrium curves $T_e=T_i$ are available in the paper \cite{LeeMore:PhysFluids:1984}. The Lee-More results are available for the density 2.5~g/cm$^3$. We did not perform calculation for that particular density, but our data for the surrounding densities 2.35~g/cm$^3$ and 2.70~g/cm$^3$ are available. 

The results on the static electrical conductivity are shown in Fig.~\ref{Fig:LeeMore}a. At high temperatures the Lee-More model yields the results consistent with the Spitzer theory. At temperature lowering the Spitzer results decrease rapidly. The Lee-More model demonstrates the slower decrease of the electrical conductivity. Though the treatment of the non-ideal plasma within the Lee-More model is approximate, this behavior, different from the Spitzer results is the important feature of this model. The Lee-More transport coefficients from the phase shift calculation are consistent with the results obtained using the Coulomb potential with cutoffs (Fig.~\ref{Fig:LeeMore}). At low temperatures the Lee-More model reduces to the Drude model with the electron-ion relaxation time $\propto T_i^{-1}$. So when the temperature increases, $\sigma_{1_\mathrm{DC}}$ decreases faster than our results, approximately described by expression (\ref{Eq:ResigmaDC_approx}). The inconsistency of our values with those of Lee-More may also be explained by the inaccurate coefficient used by Lee and More in the Bloch-Gr\"uneisen expression. The two branches of the Lee-More model, low-temperature and high-temperature, yield the minimum of the static electrical conductivity. In this region some approximation is used for the electron-ion relaxation time, and, therefore the position of the minimum is inaccurate. However, the very existence of this minimum and the construction of the wide range model are the significant advantages of the Lee-More work.

The results on the thermal conductivity are shown in Fig.~\ref{Fig:LeeMore}b. Again at high temperatures the Lee-More model is consistent with the Spitzer model. At lower temperatures the two methods of Lee-More yield consistent results slightly different from the Spitzer model. At low temperatures the thermal conductivity increases as the temperature grows. The rate of this increase is smaller than that of our \textit{ab initio} data. Again this is explained by the electron-ion relaxation time $\propto T_i^{-1}$. In fact this behavior of the Lee-More model is similar to the constant behavior demonstrated by the Anisimov model (see Fig.~\ref{Fig:Anisimov_Inogamov}a).

\subsubsection{The Apfelbaum model}

We have also compared our results with the Apfelbaum model \cite{Apfelbaum:HTHP:2008, Apfelbaum:CPP:2013}.

As well as the Lee-More model the Apfelbaum calculation of transport properties is based on the kinetic Boltzmann equation with the collision term in the approximation of relaxation time. 

Unlike the Lee-More model, where electron-electron collisions are neglected, in the Apfelbaum model they are taken into account by the introduction of the so-called generalized Spitzer factors \cite{Apfelbaum:CPP:2013}.

The Lee-More model considers some average ionization state to calculate the electron-ion relaxation time. This average ionization state is calculated within the Tomas-Fermi approximation. Desjarlais \cite{Desjarlais:CPP:2001} uses the blend of the Tomas-Fermi model and the non-ideal Saha model to calculate average ionization state. The further improvement is to calculate the chemical composition of plasma. This was implemented in the number of papers \cite{Redmer:PRE:1999, Kuhlbrodt:PRE:2000, Apfelbaum:CPP:2013}. The calculation of the chemical composition of plasma instead of the average ionization state approximation increases the reliability of the model.

In the Apfelbaum approach the chemical composition of plasma is calculated using the generalized chemical model. The total inverse electron-ion relaxation time is calculated as the sum of the inverse times of the electron relaxation on the particular sort of ions \cite{Apfelbaum:CPP:2013}. 

Like in the Lee-More model, the Coulomb potential with cutoffs is used to calculate the time of collision of electrons with the particular sort of ions, however, no corrections for degeneracy and strong coupling are introduced in paper \cite{Apfelbaum:CPP:2013}. Since that the Apfelbaum model may be used only for non-degenerate plasma with small non-ideality. At normal density the model is applicable only for $T>100$~kK.

\begin{figure}
	\includegraphics[width=0.95\columnwidth]{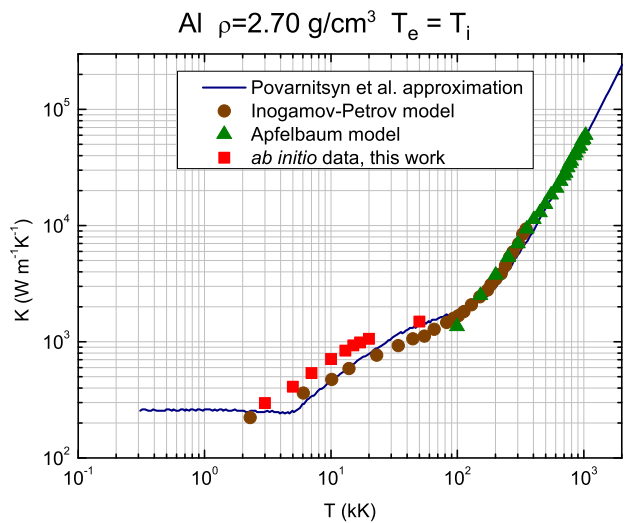}
	\caption{(Color online) The wide-range temperature dependence of the thermal conductivity. Squares---\textit{ab initio} data of this work. Circles---the Inogamov-Petrov model \cite{Inogamov:JETP:2010}. Triangles---the Apfelbaum model \cite{Apfelbaum:HTHP:2008}. Solid line---the wide-range approximation of Povarintsyn et al. \cite{Povarnitsyn:ASS:2012}}
	\label{Fig:Apfelbaum_WR}
\end{figure}

Both \textit{ab initio} results of this work and the calculation by the Apfelbaum model are presented in Fig.~\ref{Fig:Apfelbaum_WR}; the dependences conjugate rather well (by the factor of 1.1).

\subsubsection{The wide-range Povarnitsyn expression}

The wide-range approximations of the transport and optical properties are commonly built to combine the results of different models in the form applicable for practical use. The Povarnitsyn et al. model \cite{Povarnitsyn:ASS:2012} is the example of such an approximation.

At temperatures, significantly lower than the Fermi one, the thermal conductivity in the Povarnitsyn model is calculated according to the Drude expression (\ref{Eq:K_Drude}). At temperatures above melting, the relaxation time $\tau_e$ is calculated according to the common expression $1/\tau_e=AT_i+BT_e^2$. This relaxation time decreases if $T_i$ and $T_e$ grow. However, some minimal relaxation time $\tau_\mathrm{min}=Cr_0/v_F$ is introduced \cite{Eidmann:PRE:2000}. Here $v_F$ is the Fermi velocity, $r_0$ is the average interatomic distance, $C$ is the dimensionless constant. If $1/(AT_i+BT_e^2)<\tau_\mathrm{min}$ the relaxation time is set to $\tau_\mathrm{min}$ which is almost independent of the temperatures at $T_i,T_e<50$~kK.

At the temperatures comparable with the Fermi one the Drude dependence is damped exponentially and the Spitzer expression is smoothly switched on.

The Povarnitsyn model \cite{Povarnitsyn:ASS:2012} contains parameters which may be adjusted to fit the results of different calculations. The results of the Povarnitsyn model, together with discussed already Inogamov-Petrov, Apfelbaum models and \textit{ab initio} data of this work are presented in Fig.~\ref{Fig:Apfelbaum_WR}.

At rather low temperatures in the equilibrium case $AT\gg BT^2$ and the Povarnitsyn model yields the constant behavior, characteristic of the crystalline phase (Fig.~\ref{Fig:Apfelbaum_WR}). This is not supported by our results and was discussed previously in connection with the Anisimov model. However, already starting from the temperature $T=5$~kK the relaxation time is kept fixed at $\tau_\mathrm{min}$ and, therefore, the thermal conductivity exhibits $\propto T$ behavior. Our approximation (\ref{Eq:L22_approx}) in the equilibrium case yields $\propto T^{0.75}$ behavior. Nevertheless, the linear growth of the Povarnitsyn model is better than the constant behavior. At higher temperatures we see the smooth transition from the metallic Drude behavior to the Spitzer $\propto T^{5/2}$ behavior. The high temperature branch of the Povarnitsyn approximation reproduces well the Apfelbaum results \cite{Apfelbaum:CPP:2013}.

In the double logarithmic scale the Povarnitsyn wide-range model well reproduces the results of all works depicted in Fig.~\ref{Fig:Apfelbaum_WR}.

\section{CONCLUSIONS}

\begin{enumerate}
\item We have performed \textit{ab initio} calculations of optical properties, static electrical and thermal conductivities of liquid aluminum in the two-temperature regime.
\item Based on the results of \textit{ab initio} calculations we have built the semiempirical approximation of static electrical and thermal conductivities, and optical properties (\ref{Eq:ResigmaDC_approx})-(\ref{Eq:L22_approx}),(\ref{Eq:Resigma_approx}). The approximation is valid for liquid aluminum at normal density $\rho=2.70$~g/cm$^3$, 3~kK~$\leq T_i \leq T_e \leq 20$~kK.
\item We have found out that our results are well described by the Drude model if $\tau\propto T_i^{-0.25}$ expression is used for the relaxation time (Eq.~\ref{Eq:tau_approx}).
\item The models we have considered are all reduced at low temperatures to the Drude model with different expressions for the relaxation time. Ivanov, Anisimov and Lee-More models use the crystalline-like $\tau\propto T_i^{-1}$ expressions which are not consistent with our results. Inogamov-Petrov and Povarnitsyn models use the expressions that decline slower than $\propto T_i^{-1}$ with the temperature rise in agreement with our \textit{ab initio} data.
\end{enumerate}

\section*{ACKNOWLEDGEMENTS}
We thank E.M. Apfelbaum for fruitful discussions on the matter of this work. We thank M.E. Povarnitsyn for supplying us with the program module for the calculation of the thermal conductivity. We thank N.A. Inogamov and Yu.V. Petrov for supplying us with the data calculated according to their model.

The work was carried out with the financial support of SAEC “Rosatom” and  Helmholtz Association, FAIR-Russia Research Center Grants (2011-2014); and the Russian Foundation for Basic Research, grants 13-08-12248 and 14-08-31450.

%\bibliography{E:/KnyazevBibliography/KnyazevBibliography}

\begin{thebibliography}{55}%
\makeatletter
\providecommand \@ifxundefined [1]{%
 \@ifx{#1\undefined}
}%
\providecommand \@ifnum [1]{%
 \ifnum #1\expandafter \@firstoftwo
 \else \expandafter \@secondoftwo
 \fi
}%
\providecommand \@ifx [1]{%
 \ifx #1\expandafter \@firstoftwo
 \else \expandafter \@secondoftwo
 \fi
}%
\providecommand \natexlab [1]{#1}%
\providecommand \enquote  [1]{``#1''}%
\providecommand \bibnamefont  [1]{#1}%
\providecommand \bibfnamefont [1]{#1}%
\providecommand \citenamefont [1]{#1}%
\providecommand \href@noop [0]{\@secondoftwo}%
\providecommand \href [0]{\begingroup \@sanitize@url \@href}%
\providecommand \@href[1]{\@@startlink{#1}\@@href}%
\providecommand \@@href[1]{\endgroup#1\@@endlink}%
\providecommand \@sanitize@url [0]{\catcode `\\12\catcode `\$12\catcode
  `\&12\catcode `\#12\catcode `\^12\catcode `\_12\catcode `\%12\relax}%
\providecommand \@@startlink[1]{}%
\providecommand \@@endlink[0]{}%
\providecommand \url  [0]{\begingroup\@sanitize@url \@url }%
\providecommand \@url [1]{\endgroup\@href {#1}{\urlprefix }}%
\providecommand \urlprefix  [0]{URL }%
\providecommand \Eprint [0]{\href }%
\providecommand \doibase [0]{http://dx.doi.org/}%
\providecommand \selectlanguage [0]{\@gobble}%
\providecommand \bibinfo  [0]{\@secondoftwo}%
\providecommand \bibfield  [0]{\@secondoftwo}%
\providecommand \translation [1]{[#1]}%
\providecommand \BibitemOpen [0]{}%
\providecommand \bibitemStop [0]{}%
\providecommand \bibitemNoStop [0]{.\EOS\space}%
\providecommand \EOS [0]{\spacefactor3000\relax}%
\providecommand \BibitemShut  [1]{\csname bibitem#1\endcsname}%
\let\auto@bib@innerbib\@empty
%</preamble>
\bibitem [{\citenamefont {Chen}\ \emph {et~al.}(2013)\citenamefont {Chen},
  \citenamefont {Holst}, \citenamefont {Kirkwood}, \citenamefont {Sametoglu},
  \citenamefont {Reid}, \citenamefont {Tsui}, \citenamefont {Recoules},\ and\
  \citenamefont {Ng}}]{Chen:PRL:2013}%
  \BibitemOpen
  \bibfield  {author} {\bibinfo {author} {\bibfnamefont {Z.}~\bibnamefont
  {Chen}}, \bibinfo {author} {\bibfnamefont {B.}~\bibnamefont {Holst}},
  \bibinfo {author} {\bibfnamefont {S.~E.}\ \bibnamefont {Kirkwood}}, \bibinfo
  {author} {\bibfnamefont {V.}~\bibnamefont {Sametoglu}}, \bibinfo {author}
  {\bibfnamefont {M.}~\bibnamefont {Reid}}, \bibinfo {author} {\bibfnamefont
  {Y.~Y.}\ \bibnamefont {Tsui}}, \bibinfo {author} {\bibfnamefont
  {V.}~\bibnamefont {Recoules}}, \ and\ \bibinfo {author} {\bibfnamefont
  {A.}~\bibnamefont {Ng}},\ }\href@noop {} {\bibfield  {journal} {\bibinfo
  {journal} {Phys. Rev. Lett.}\ }\textbf {\bibinfo {volume} {110}},\ \bibinfo
  {pages} {135001} (\bibinfo {year} {2013})}\BibitemShut {NoStop}%
\bibitem [{\citenamefont {Widmann}\ \emph {et~al.}(2004)\citenamefont
  {Widmann}, \citenamefont {Ao}, \citenamefont {Foord}, \citenamefont {Price},
  \citenamefont {Ellis}, \citenamefont {Springer},\ and\ \citenamefont
  {Ng}}]{Widmann:PRL:2004}%
  \BibitemOpen
  \bibfield  {author} {\bibinfo {author} {\bibfnamefont {K.}~\bibnamefont
  {Widmann}}, \bibinfo {author} {\bibfnamefont {T.}~\bibnamefont {Ao}},
  \bibinfo {author} {\bibfnamefont {M.~E.}\ \bibnamefont {Foord}}, \bibinfo
  {author} {\bibfnamefont {D.~F.}\ \bibnamefont {Price}}, \bibinfo {author}
  {\bibfnamefont {A.~D.}\ \bibnamefont {Ellis}}, \bibinfo {author}
  {\bibfnamefont {P.~T.}\ \bibnamefont {Springer}}, \ and\ \bibinfo {author}
  {\bibfnamefont {A.}~\bibnamefont {Ng}},\ }\href@noop {} {\bibfield  {journal}
  {\bibinfo  {journal} {Phys. Rev. Lett.}\ }\textbf {\bibinfo {volume} {92}},\
  \bibinfo {pages} {125002} (\bibinfo {year} {2004})}\BibitemShut {NoStop}%
\bibitem [{\citenamefont {Chimier}\ \emph {et~al.}(2011)\citenamefont
  {Chimier}, \citenamefont {Ut\'eza}, \citenamefont {Sanner}, \citenamefont
  {Sentis}, \citenamefont {Itina}, \citenamefont {Lassonde}, \citenamefont
  {L\'egar\'e}, \citenamefont {Vidal},\ and\ \citenamefont
  {Kieffer}}]{Chimier:PRB:2011}%
  \BibitemOpen
  \bibfield  {author} {\bibinfo {author} {\bibfnamefont {B.}~\bibnamefont
  {Chimier}}, \bibinfo {author} {\bibfnamefont {O.}~\bibnamefont {Ut\'eza}},
  \bibinfo {author} {\bibfnamefont {N.}~\bibnamefont {Sanner}}, \bibinfo
  {author} {\bibfnamefont {M.}~\bibnamefont {Sentis}}, \bibinfo {author}
  {\bibfnamefont {T.}~\bibnamefont {Itina}}, \bibinfo {author} {\bibfnamefont
  {P.}~\bibnamefont {Lassonde}}, \bibinfo {author} {\bibfnamefont
  {F.}~\bibnamefont {L\'egar\'e}}, \bibinfo {author} {\bibfnamefont
  {F.}~\bibnamefont {Vidal}}, \ and\ \bibinfo {author} {\bibfnamefont {J.~C.}\
  \bibnamefont {Kieffer}},\ }\href@noop {} {\bibfield  {journal} {\bibinfo
  {journal} {Phys. Rev. B}\ }\textbf {\bibinfo {volume} {84}},\ \bibinfo
  {pages} {094104} (\bibinfo {year} {2011})}\BibitemShut {NoStop}%
\bibitem [{\citenamefont {Veysman}\ \emph {et~al.}(2008)\citenamefont
  {Veysman}, \citenamefont {Agranat}, \citenamefont {Andreev}, \citenamefont
  {Ashitkov}, \citenamefont {Fortov}, \citenamefont {Khishchenko},
  \citenamefont {Kostenko}, \citenamefont {Levashov}, \citenamefont
  {Ovchinnikov},\ and\ \citenamefont {Sitnikov}}]{Veysman:JPB:2008}%
  \BibitemOpen
  \bibfield  {author} {\bibinfo {author} {\bibfnamefont {M.~E.}\ \bibnamefont
  {Veysman}}, \bibinfo {author} {\bibfnamefont {M.~B.}\ \bibnamefont
  {Agranat}}, \bibinfo {author} {\bibfnamefont {N.~E.}\ \bibnamefont
  {Andreev}}, \bibinfo {author} {\bibfnamefont {S.~I.}\ \bibnamefont
  {Ashitkov}}, \bibinfo {author} {\bibfnamefont {V.~E.}\ \bibnamefont
  {Fortov}}, \bibinfo {author} {\bibfnamefont {K.~V.}\ \bibnamefont
  {Khishchenko}}, \bibinfo {author} {\bibfnamefont {O.~F.}\ \bibnamefont
  {Kostenko}}, \bibinfo {author} {\bibfnamefont {P.~R.}\ \bibnamefont
  {Levashov}}, \bibinfo {author} {\bibfnamefont {A.~V.}\ \bibnamefont
  {Ovchinnikov}}, \ and\ \bibinfo {author} {\bibfnamefont {D.~S.}\ \bibnamefont
  {Sitnikov}},\ }\href@noop {} {\bibfield  {journal} {\bibinfo  {journal} {J.
  Phys. B: At. Mol. Opt. Phys.}\ }\textbf {\bibinfo {volume} {41}},\ \bibinfo
  {pages} {125704} (\bibinfo {year} {2008})}\BibitemShut {NoStop}%
\bibitem [{\citenamefont {Povarnitsyn}\ \emph
  {et~al.}(2012{\natexlab{a}})\citenamefont {Povarnitsyn}, \citenamefont
  {Andreev}, \citenamefont {Apfelbaum}, \citenamefont {Itina}, \citenamefont
  {Khishchenko}, \citenamefont {Kostenko}, \citenamefont {Levashov},\ and\
  \citenamefont {Veysman}}]{Povarnitsyn:ASS:2012}%
  \BibitemOpen
  \bibfield  {author} {\bibinfo {author} {\bibfnamefont {M.~E.}\ \bibnamefont
  {Povarnitsyn}}, \bibinfo {author} {\bibfnamefont {N.~E.}\ \bibnamefont
  {Andreev}}, \bibinfo {author} {\bibfnamefont {E.~M.}\ \bibnamefont
  {Apfelbaum}}, \bibinfo {author} {\bibfnamefont {T.~E.}\ \bibnamefont
  {Itina}}, \bibinfo {author} {\bibfnamefont {K.~V.}\ \bibnamefont
  {Khishchenko}}, \bibinfo {author} {\bibfnamefont {O.~F.}\ \bibnamefont
  {Kostenko}}, \bibinfo {author} {\bibfnamefont {P.~R.}\ \bibnamefont
  {Levashov}}, \ and\ \bibinfo {author} {\bibfnamefont {M.}~\bibnamefont
  {Veysman}},\ }\href@noop {} {\bibfield  {journal} {\bibinfo  {journal}
  {Applied Surface Science}\ }\textbf {\bibinfo {volume} {258}},\ \bibinfo
  {pages} {9480} (\bibinfo {year} {2012}{\natexlab{a}})}\BibitemShut {NoStop}%
\bibitem [{\citenamefont {Ivanov}\ and\ \citenamefont
  {Zhigilei}(2003)}]{Ivanov:PRB:2003}%
  \BibitemOpen
  \bibfield  {author} {\bibinfo {author} {\bibfnamefont {D.~S.}\ \bibnamefont
  {Ivanov}}\ and\ \bibinfo {author} {\bibfnamefont {L.~V.}\ \bibnamefont
  {Zhigilei}},\ }\href@noop {} {\bibfield  {journal} {\bibinfo  {journal}
  {Phys. Rev. B}\ }\textbf {\bibinfo {volume} {68}},\ \bibinfo {pages} {064114}
  (\bibinfo {year} {2003})}\BibitemShut {NoStop}%
\bibitem [{\citenamefont {Mazevet}\ \emph {et~al.}(2005)\citenamefont
  {Mazevet}, \citenamefont {Cl\'erouin}, \citenamefont {Recoules},
  \citenamefont {Anglade},\ and\ \citenamefont {Zerah}}]{Mazevet:PRL:2005}%
  \BibitemOpen
  \bibfield  {author} {\bibinfo {author} {\bibfnamefont {S.}~\bibnamefont
  {Mazevet}}, \bibinfo {author} {\bibfnamefont {J.}~\bibnamefont {Cl\'erouin}},
  \bibinfo {author} {\bibfnamefont {V.}~\bibnamefont {Recoules}}, \bibinfo
  {author} {\bibfnamefont {P.~M.}\ \bibnamefont {Anglade}}, \ and\ \bibinfo
  {author} {\bibfnamefont {G.}~\bibnamefont {Zerah}},\ }\href@noop {}
  {\bibfield  {journal} {\bibinfo  {journal} {Phys. Rev. Lett.}\ }\textbf
  {\bibinfo {volume} {95}},\ \bibinfo {pages} {085002} (\bibinfo {year}
  {2005})}\BibitemShut {NoStop}%
\bibitem [{\citenamefont {Kirkwood}\ \emph {et~al.}(2009)\citenamefont
  {Kirkwood}, \citenamefont {Tsui}, \citenamefont {Fedosejevs}, \citenamefont
  {Brantov},\ and\ \citenamefont {Bychenkov}}]{Kirkwood:PRB:2009}%
  \BibitemOpen
  \bibfield  {author} {\bibinfo {author} {\bibfnamefont {S.~E.}\ \bibnamefont
  {Kirkwood}}, \bibinfo {author} {\bibfnamefont {Y.~Y.}\ \bibnamefont {Tsui}},
  \bibinfo {author} {\bibfnamefont {R.}~\bibnamefont {Fedosejevs}}, \bibinfo
  {author} {\bibfnamefont {A.~V.}\ \bibnamefont {Brantov}}, \ and\ \bibinfo
  {author} {\bibfnamefont {V.~Y.}\ \bibnamefont {Bychenkov}},\ }\href@noop {}
  {\bibfield  {journal} {\bibinfo  {journal} {Phys. Rev. B}\ }\textbf {\bibinfo
  {volume} {79}},\ \bibinfo {pages} {144120} (\bibinfo {year}
  {2009})}\BibitemShut {NoStop}%
\bibitem [{\citenamefont {Loboda}\ \emph {et~al.}(2011)\citenamefont {Loboda},
  \citenamefont {Smirnov}, \citenamefont {Shadrin},\ and\ \citenamefont
  {Karlykhanov}}]{Loboda:HEDP:2011}%
  \BibitemOpen
  \bibfield  {author} {\bibinfo {author} {\bibfnamefont {P.~A.}\ \bibnamefont
  {Loboda}}, \bibinfo {author} {\bibfnamefont {N.~A.}\ \bibnamefont {Smirnov}},
  \bibinfo {author} {\bibfnamefont {A.~A.}\ \bibnamefont {Shadrin}}, \ and\
  \bibinfo {author} {\bibfnamefont {N.~G.}\ \bibnamefont {Karlykhanov}},\
  }\href@noop {} {\bibfield  {journal} {\bibinfo  {journal} {High Energy
  Density Physics}\ }\textbf {\bibinfo {volume} {7}},\ \bibinfo {pages} {361}
  (\bibinfo {year} {2011})}\BibitemShut {NoStop}%
\bibitem [{\citenamefont {Dharma-wardana}\ and\ \citenamefont
  {Murillo}(2008)}]{Dharma-wardana:PRE:2008}%
  \BibitemOpen
  \bibfield  {author} {\bibinfo {author} {\bibfnamefont {M.~W.~C.}\
  \bibnamefont {Dharma-wardana}}\ and\ \bibinfo {author} {\bibfnamefont
  {M.~S.}\ \bibnamefont {Murillo}},\ }\href@noop {} {\bibfield  {journal}
  {\bibinfo  {journal} {Phys. Rev. E}\ }\textbf {\bibinfo {volume} {77}},\
  \bibinfo {pages} {026401} (\bibinfo {year} {2008})}\BibitemShut {NoStop}%
\bibitem [{\citenamefont {Recoules}\ \emph {et~al.}(2006)\citenamefont
  {Recoules}, \citenamefont {Cl\'erouin}, \citenamefont {Z\'erah},
  \citenamefont {Anglade},\ and\ \citenamefont {Mazevet}}]{Recoules:PRL:2006}%
  \BibitemOpen
  \bibfield  {author} {\bibinfo {author} {\bibfnamefont {V.}~\bibnamefont
  {Recoules}}, \bibinfo {author} {\bibfnamefont {J.}~\bibnamefont
  {Cl\'erouin}}, \bibinfo {author} {\bibfnamefont {G.}~\bibnamefont {Z\'erah}},
  \bibinfo {author} {\bibfnamefont {P.~M.}\ \bibnamefont {Anglade}}, \ and\
  \bibinfo {author} {\bibfnamefont {S.}~\bibnamefont {Mazevet}},\ }\href@noop
  {} {\bibfield  {journal} {\bibinfo  {journal} {Phys. Rev. Lett.}\ }\textbf
  {\bibinfo {volume} {96}},\ \bibinfo {pages} {055503} (\bibinfo {year}
  {2006})}\BibitemShut {NoStop}%
\bibitem [{\citenamefont {Ernstorfer}\ \emph {et~al.}(2009)\citenamefont
  {Ernstorfer}, \citenamefont {Harb}, \citenamefont {Hebeisen}, \citenamefont
  {Sciani}, \citenamefont {Dartigalongue},\ and\ \citenamefont
  {Dwayne~Miller}}]{Ernstorfer:Science:2009}%
  \BibitemOpen
  \bibfield  {author} {\bibinfo {author} {\bibfnamefont {R.}~\bibnamefont
  {Ernstorfer}}, \bibinfo {author} {\bibfnamefont {M.}~\bibnamefont {Harb}},
  \bibinfo {author} {\bibfnamefont {C.~T.}\ \bibnamefont {Hebeisen}}, \bibinfo
  {author} {\bibfnamefont {G.}~\bibnamefont {Sciani}}, \bibinfo {author}
  {\bibfnamefont {T.}~\bibnamefont {Dartigalongue}}, \ and\ \bibinfo {author}
  {\bibfnamefont {R.~J.}\ \bibnamefont {Dwayne~Miller}},\ }\href@noop {}
  {\bibfield  {journal} {\bibinfo  {journal} {Science}\ }\textbf {\bibinfo
  {volume} {323}},\ \bibinfo {pages} {1033} (\bibinfo {year}
  {2009})}\BibitemShut {NoStop}%
\bibitem [{\citenamefont {Lin}\ \emph {et~al.}(2008)\citenamefont {Lin},
  \citenamefont {Zhigilei},\ and\ \citenamefont {Celli}}]{Lin:PRB:2008}%
  \BibitemOpen
  \bibfield  {author} {\bibinfo {author} {\bibfnamefont {{\relax
  Zh}.}~\bibnamefont {Lin}}, \bibinfo {author} {\bibfnamefont {L.~V.}\
  \bibnamefont {Zhigilei}}, \ and\ \bibinfo {author} {\bibfnamefont
  {V.}~\bibnamefont {Celli}},\ }\href@noop {} {\bibfield  {journal} {\bibinfo
  {journal} {Phys. Rev. B}\ }\textbf {\bibinfo {volume} {77}},\ \bibinfo
  {pages} {075133} (\bibinfo {year} {2008})}\BibitemShut {NoStop}%
\bibitem [{\citenamefont {Stegailov}(2010)}]{Stegailov:CPP:2010}%
  \BibitemOpen
  \bibfield  {author} {\bibinfo {author} {\bibfnamefont {V.~V.}\ \bibnamefont
  {Stegailov}},\ }\href@noop {} {\bibfield  {journal} {\bibinfo  {journal}
  {Contrib. Plasma Phys.}\ }\textbf {\bibinfo {volume} {50}},\ \bibinfo {pages}
  {31} (\bibinfo {year} {2010})}\BibitemShut {NoStop}%
\bibitem [{\citenamefont {Dharma-wardana}\ and\ \citenamefont
  {Perrot}(1998)}]{Dharma-wardana:PRE:1998}%
  \BibitemOpen
  \bibfield  {author} {\bibinfo {author} {\bibfnamefont {M.~W.~C.}\
  \bibnamefont {Dharma-wardana}}\ and\ \bibinfo {author} {\bibfnamefont
  {F.}~\bibnamefont {Perrot}},\ }\href@noop {} {\bibfield  {journal} {\bibinfo
  {journal} {Phys. Rev. E}\ }\textbf {\bibinfo {volume} {58}},\ \bibinfo
  {pages} {3705} (\bibinfo {year} {1998})}\BibitemShut {NoStop}%
\bibitem [{\citenamefont {Levashov}\ \emph {et~al.}(2010)\citenamefont
  {Levashov}, \citenamefont {Sin'ko}, \citenamefont {Smirnov}, \citenamefont
  {Minakov}, \citenamefont {Shemyakin},\ and\ \citenamefont
  {Khishchenko}}]{Levashov:JPCM:2010}%
  \BibitemOpen
  \bibfield  {author} {\bibinfo {author} {\bibfnamefont {P.~R.}\ \bibnamefont
  {Levashov}}, \bibinfo {author} {\bibfnamefont {G.~V.}\ \bibnamefont
  {Sin'ko}}, \bibinfo {author} {\bibfnamefont {N.~A.}\ \bibnamefont {Smirnov}},
  \bibinfo {author} {\bibfnamefont {D.~V.}\ \bibnamefont {Minakov}}, \bibinfo
  {author} {\bibfnamefont {O.~P.}\ \bibnamefont {Shemyakin}}, \ and\ \bibinfo
  {author} {\bibfnamefont {K.~V.}\ \bibnamefont {Khishchenko}},\ }\href@noop {}
  {\bibfield  {journal} {\bibinfo  {journal} {J. Phys.: Condens. Matter}\
  }\textbf {\bibinfo {volume} {22}},\ \bibinfo {pages} {505501} (\bibinfo
  {year} {2010})}\BibitemShut {NoStop}%
\bibitem [{\citenamefont {Sin'ko}\ \emph {et~al.}(2013)\citenamefont {Sin'ko},
  \citenamefont {Smirnov}, \citenamefont {Ovechkin}, \citenamefont {Levashov},\
  and\ \citenamefont {Khishchenko}}]{Sinko:HEDP:2013}%
  \BibitemOpen
  \bibfield  {author} {\bibinfo {author} {\bibfnamefont {G.~V.}\ \bibnamefont
  {Sin'ko}}, \bibinfo {author} {\bibfnamefont {N.~A.}\ \bibnamefont {Smirnov}},
  \bibinfo {author} {\bibfnamefont {A.~A.}\ \bibnamefont {Ovechkin}}, \bibinfo
  {author} {\bibfnamefont {P.~R.}\ \bibnamefont {Levashov}}, \ and\ \bibinfo
  {author} {\bibfnamefont {K.~V.}\ \bibnamefont {Khishchenko}},\ }\href@noop {}
  {\bibfield  {journal} {\bibinfo  {journal} {High Energy Density Physics}\
  }\textbf {\bibinfo {volume} {9}},\ \bibinfo {pages} {309} (\bibinfo {year}
  {2013})}\BibitemShut {NoStop}%
\bibitem [{\citenamefont {Lee}\ and\ \citenamefont
  {More}(1984)}]{LeeMore:PhysFluids:1984}%
  \BibitemOpen
  \bibfield  {author} {\bibinfo {author} {\bibfnamefont {Y.~T.}\ \bibnamefont
  {Lee}}\ and\ \bibinfo {author} {\bibfnamefont {R.~M.}\ \bibnamefont {More}},\
  }\href@noop {} {\bibfield  {journal} {\bibinfo  {journal} {Phys. Fluids}\
  }\textbf {\bibinfo {volume} {27}},\ \bibinfo {pages} {1273} (\bibinfo {year}
  {1984})}\BibitemShut {NoStop}%
\bibitem [{\citenamefont {Eidmann}\ \emph {et~al.}(2000)\citenamefont
  {Eidmann}, \citenamefont {{Meyer-ter-Vehn}}, \citenamefont {Schlegel},\ and\
  \citenamefont {H\"uller}}]{Eidmann:PRE:2000}%
  \BibitemOpen
  \bibfield  {author} {\bibinfo {author} {\bibfnamefont {K.}~\bibnamefont
  {Eidmann}}, \bibinfo {author} {\bibfnamefont {J.}~\bibnamefont
  {{Meyer-ter-Vehn}}}, \bibinfo {author} {\bibfnamefont {T.}~\bibnamefont
  {Schlegel}}, \ and\ \bibinfo {author} {\bibfnamefont {S.}~\bibnamefont
  {H\"uller}},\ }\href@noop {} {\bibfield  {journal} {\bibinfo  {journal}
  {Phys. Rev. E}\ }\textbf {\bibinfo {volume} {62}},\ \bibinfo {pages} {1202}
  (\bibinfo {year} {2000})}\BibitemShut {NoStop}%
\bibitem [{\citenamefont {Spitzer}\ and\ \citenamefont
  {H\"arm}(1953)}]{Spitzer:PR:1953}%
  \BibitemOpen
  \bibfield  {author} {\bibinfo {author} {\bibfnamefont {L.}~\bibnamefont
  {Spitzer}}\ and\ \bibinfo {author} {\bibfnamefont {R.}~\bibnamefont
  {H\"arm}},\ }\href@noop {} {\bibfield  {journal} {\bibinfo  {journal} {Phys.
  Rev.}\ }\textbf {\bibinfo {volume} {89}},\ \bibinfo {pages} {977} (\bibinfo
  {year} {1953})}\BibitemShut {NoStop}%
\bibitem [{\citenamefont {Apfelbaum}(2013)}]{Apfelbaum:CPP:2013}%
  \BibitemOpen
  \bibfield  {author} {\bibinfo {author} {\bibfnamefont {E.~M.}\ \bibnamefont
  {Apfelbaum}},\ }\href@noop {} {\bibfield  {journal} {\bibinfo  {journal}
  {Contrib. Plasma Phys.}\ }\textbf {\bibinfo {volume} {53}},\ \bibinfo {pages}
  {317} (\bibinfo {year} {2013})}\BibitemShut {NoStop}%
\bibitem [{\citenamefont {Johnson}\ \emph {et~al.}(2006)\citenamefont
  {Johnson}, \citenamefont {Guet},\ and\ \citenamefont
  {Bertsch}}]{Johnson:JQSRP:2006}%
  \BibitemOpen
  \bibfield  {author} {\bibinfo {author} {\bibfnamefont {W.}~\bibnamefont
  {Johnson}}, \bibinfo {author} {\bibfnamefont {C.}~\bibnamefont {Guet}}, \
  and\ \bibinfo {author} {\bibfnamefont {G.}~\bibnamefont {Bertsch}},\
  }\href@noop {} {\bibfield  {journal} {\bibinfo  {journal} {Journal of
  Quantitative Spectroscopy \& Radiative Transfer}\ }\textbf {\bibinfo {volume}
  {99}},\ \bibinfo {pages} {327} (\bibinfo {year} {2006})}\BibitemShut
  {NoStop}%
\bibitem [{\citenamefont {Starrett}\ \emph {et~al.}(2012)\citenamefont
  {Starrett}, \citenamefont {Cl\'erouin}, \citenamefont {Recoules},
  \citenamefont {Kress}, \citenamefont {Collins},\ and\ \citenamefont
  {Hanson}}]{Starrett:PhysPlasmas:2012}%
  \BibitemOpen
  \bibfield  {author} {\bibinfo {author} {\bibfnamefont {C.~E.}\ \bibnamefont
  {Starrett}}, \bibinfo {author} {\bibfnamefont {J.}~\bibnamefont
  {Cl\'erouin}}, \bibinfo {author} {\bibfnamefont {V.}~\bibnamefont
  {Recoules}}, \bibinfo {author} {\bibfnamefont {J.~D.}\ \bibnamefont {Kress}},
  \bibinfo {author} {\bibfnamefont {L.~A.}\ \bibnamefont {Collins}}, \ and\
  \bibinfo {author} {\bibfnamefont {D.~E.}\ \bibnamefont {Hanson}},\
  }\href@noop {} {\bibfield  {journal} {\bibinfo  {journal} {Phys. Plasmas}\
  }\textbf {\bibinfo {volume} {19}},\ \bibinfo {pages} {102709} (\bibinfo
  {year} {2012})}\BibitemShut {NoStop}%
\bibitem [{\citenamefont {Johnson}(2009)}]{Johnson:HEDP:2009}%
  \BibitemOpen
  \bibfield  {author} {\bibinfo {author} {\bibfnamefont {W.~D.}\ \bibnamefont
  {Johnson}},\ }\href@noop {} {\bibfield  {journal} {\bibinfo  {journal} {High
  Energy Density Physics}\ }\textbf {\bibinfo {volume} {5}},\ \bibinfo {pages}
  {61} (\bibinfo {year} {2009})}\BibitemShut {NoStop}%
\bibitem [{\citenamefont {Apfelbaum}(2006)}]{Apfelbaum:JPhysA:2006}%
  \BibitemOpen
  \bibfield  {author} {\bibinfo {author} {\bibfnamefont {E.~M.}\ \bibnamefont
  {Apfelbaum}},\ }\href@noop {} {\bibfield  {journal} {\bibinfo  {journal} {J.
  Phys. A: Math. Gen.}\ }\textbf {\bibinfo {volume} {39}},\ \bibinfo {pages}
  {4407} (\bibinfo {year} {2006})}\BibitemShut {NoStop}%
\bibitem [{\citenamefont {Dharma-wardana}(2006)}]{Dharma-wardana:PRE:2006}%
  \BibitemOpen
  \bibfield  {author} {\bibinfo {author} {\bibfnamefont {M.~W.~C.}\
  \bibnamefont {Dharma-wardana}},\ }\href@noop {} {\bibfield  {journal}
  {\bibinfo  {journal} {Phys. Rev. E}\ }\textbf {\bibinfo {volume} {73}},\
  \bibinfo {pages} {036401} (\bibinfo {year} {2006})}\BibitemShut {NoStop}%
\bibitem [{\citenamefont {Fois}\ \emph {et~al.}(1989)\citenamefont {Fois},
  \citenamefont {Selloni},\ and\ \citenamefont {Parrinello}}]{Fois:PRB:1989}%
  \BibitemOpen
  \bibfield  {author} {\bibinfo {author} {\bibfnamefont {E.}~\bibnamefont
  {Fois}}, \bibinfo {author} {\bibfnamefont {A.}~\bibnamefont {Selloni}}, \
  and\ \bibinfo {author} {\bibfnamefont {M.}~\bibnamefont {Parrinello}},\
  }\href@noop {} {\bibfield  {journal} {\bibinfo  {journal} {Phys. Rev. B}\
  }\textbf {\bibinfo {volume} {39}},\ \bibinfo {pages} {4812} (\bibinfo {year}
  {1989})}\BibitemShut {NoStop}%
\bibitem [{\citenamefont {Silvestrelli}\ \emph {et~al.}(1996)\citenamefont
  {Silvestrelli}, \citenamefont {Alavi}, \citenamefont {Parrinello},\ and\
  \citenamefont {Frenkel}}]{Silvestrelli:PRB:1996}%
  \BibitemOpen
  \bibfield  {author} {\bibinfo {author} {\bibfnamefont {P.~L.}\ \bibnamefont
  {Silvestrelli}}, \bibinfo {author} {\bibfnamefont {A.}~\bibnamefont {Alavi}},
  \bibinfo {author} {\bibfnamefont {M.}~\bibnamefont {Parrinello}}, \ and\
  \bibinfo {author} {\bibfnamefont {D.}~\bibnamefont {Frenkel}},\ }\href@noop
  {} {\bibfield  {journal} {\bibinfo  {journal} {Phys. Rev. B}\ }\textbf
  {\bibinfo {volume} {53}},\ \bibinfo {pages} {12750} (\bibinfo {year}
  {1996})}\BibitemShut {NoStop}%
\bibitem [{\citenamefont {Silvestrelli}(1999)}]{Silvestrelli:PRB:1999}%
  \BibitemOpen
  \bibfield  {author} {\bibinfo {author} {\bibfnamefont {P.~L.}\ \bibnamefont
  {Silvestrelli}},\ }\href@noop {} {\bibfield  {journal} {\bibinfo  {journal}
  {Phys. Rev. B}\ }\textbf {\bibinfo {volume} {60}},\ \bibinfo {pages} {16382}
  (\bibinfo {year} {1999})}\BibitemShut {NoStop}%
\bibitem [{\citenamefont {Collins}\ \emph {et~al.}(2001)\citenamefont
  {Collins}, \citenamefont {Bickham}, \citenamefont {Kress}, \citenamefont
  {Mazevet}, \citenamefont {Lenosky}, \citenamefont {Troullier},\ and\
  \citenamefont {Windl}}]{Collins:PRB:2001}%
  \BibitemOpen
  \bibfield  {author} {\bibinfo {author} {\bibfnamefont {L.~A.}\ \bibnamefont
  {Collins}}, \bibinfo {author} {\bibfnamefont {S.~R.}\ \bibnamefont
  {Bickham}}, \bibinfo {author} {\bibfnamefont {J.~D.}\ \bibnamefont {Kress}},
  \bibinfo {author} {\bibfnamefont {S.}~\bibnamefont {Mazevet}}, \bibinfo
  {author} {\bibfnamefont {T.~J.}\ \bibnamefont {Lenosky}}, \bibinfo {author}
  {\bibfnamefont {N.~J.}\ \bibnamefont {Troullier}}, \ and\ \bibinfo {author}
  {\bibfnamefont {W.}~\bibnamefont {Windl}},\ }\href@noop {} {\bibfield
  {journal} {\bibinfo  {journal} {Phys. Rev. B}\ }\textbf {\bibinfo {volume}
  {63}},\ \bibinfo {pages} {184110} (\bibinfo {year} {2001})}\BibitemShut
  {NoStop}%
\bibitem [{\citenamefont {Desjarlais}\ \emph {et~al.}(2002)\citenamefont
  {Desjarlais}, \citenamefont {Kress},\ and\ \citenamefont
  {Collins}}]{Desjarlais:PRE:2002}%
  \BibitemOpen
  \bibfield  {author} {\bibinfo {author} {\bibfnamefont {M.~P.}\ \bibnamefont
  {Desjarlais}}, \bibinfo {author} {\bibfnamefont {J.~D.}\ \bibnamefont
  {Kress}}, \ and\ \bibinfo {author} {\bibfnamefont {L.~A.}\ \bibnamefont
  {Collins}},\ }\href@noop {} {\bibfield  {journal} {\bibinfo  {journal} {Phys.
  Rev. E}\ }\textbf {\bibinfo {volume} {66}},\ \bibinfo {pages} {025401}
  (\bibinfo {year} {2002})}\BibitemShut {NoStop}%
\bibitem [{\citenamefont {Recoules}\ and\ \citenamefont
  {Crocombette}(2005)}]{Recoules:PRB:2005}%
  \BibitemOpen
  \bibfield  {author} {\bibinfo {author} {\bibfnamefont {V.}~\bibnamefont
  {Recoules}}\ and\ \bibinfo {author} {\bibfnamefont {J.-P.}\ \bibnamefont
  {Crocombette}},\ }\href@noop {} {\bibfield  {journal} {\bibinfo  {journal}
  {Phys. Rev. B}\ }\textbf {\bibinfo {volume} {72}},\ \bibinfo {pages} {104202}
  (\bibinfo {year} {2005})}\BibitemShut {NoStop}%
\bibitem [{\citenamefont {Laudernet}\ \emph {et~al.}(2004)\citenamefont
  {Laudernet}, \citenamefont {Cl\'erouin},\ and\ \citenamefont
  {Mazevet}}]{Laudernet:PRB:2004}%
  \BibitemOpen
  \bibfield  {author} {\bibinfo {author} {\bibfnamefont {Y.}~\bibnamefont
  {Laudernet}}, \bibinfo {author} {\bibfnamefont {J.}~\bibnamefont
  {Cl\'erouin}}, \ and\ \bibinfo {author} {\bibfnamefont {S.}~\bibnamefont
  {Mazevet}},\ }\href@noop {} {\bibfield  {journal} {\bibinfo  {journal} {Phys.
  Rev. B}\ }\textbf {\bibinfo {volume} {70}},\ \bibinfo {pages} {165108}
  (\bibinfo {year} {2004})}\BibitemShut {NoStop}%
\bibitem [{\citenamefont {French}\ and\ \citenamefont
  {Redmer}(2011)}]{French:PhysPlasmas:2011}%
  \BibitemOpen
  \bibfield  {author} {\bibinfo {author} {\bibfnamefont {M.}~\bibnamefont
  {French}}\ and\ \bibinfo {author} {\bibfnamefont {R.}~\bibnamefont
  {Redmer}},\ }\href@noop {} {\bibfield  {journal} {\bibinfo  {journal} {Phys.
  Plasmas}\ }\textbf {\bibinfo {volume} {18}},\ \bibinfo {pages} {043301}
  (\bibinfo {year} {2011})}\BibitemShut {NoStop}%
\bibitem [{\citenamefont {Lambert}\ \emph {et~al.}(2011)\citenamefont
  {Lambert}, \citenamefont {Recoules}, \citenamefont {Decoster}, \citenamefont
  {Cl\'erouin},\ and\ \citenamefont {Desjarlais}}]{Lambert:PhysPlasmas:2011}%
  \BibitemOpen
  \bibfield  {author} {\bibinfo {author} {\bibfnamefont {F.}~\bibnamefont
  {Lambert}}, \bibinfo {author} {\bibfnamefont {V.}~\bibnamefont {Recoules}},
  \bibinfo {author} {\bibfnamefont {A.}~\bibnamefont {Decoster}}, \bibinfo
  {author} {\bibfnamefont {J.}~\bibnamefont {Cl\'erouin}}, \ and\ \bibinfo
  {author} {\bibfnamefont {M.}~\bibnamefont {Desjarlais}},\ }\href@noop {}
  {\bibfield  {journal} {\bibinfo  {journal} {Phys. Plasmas}\ }\textbf
  {\bibinfo {volume} {18}},\ \bibinfo {pages} {056306} (\bibinfo {year}
  {2011})}\BibitemShut {NoStop}%
\bibitem [{\citenamefont {Norman}\ \emph {et~al.}(2013)\citenamefont {Norman},
  \citenamefont {Saitov}, \citenamefont {Stegailov},\ and\ \citenamefont
  {Zhilyaev}}]{Norman:CPP:2013}%
  \BibitemOpen
  \bibfield  {author} {\bibinfo {author} {\bibfnamefont {G.}~\bibnamefont
  {Norman}}, \bibinfo {author} {\bibfnamefont {I.}~\bibnamefont {Saitov}},
  \bibinfo {author} {\bibfnamefont {V.}~\bibnamefont {Stegailov}}, \ and\
  \bibinfo {author} {\bibfnamefont {P.}~\bibnamefont {Zhilyaev}},\ }\href@noop
  {} {\bibfield  {journal} {\bibinfo  {journal} {Contrib. Plasma Phys.}\
  }\textbf {\bibinfo {volume} {53}},\ \bibinfo {pages} {300} (\bibinfo {year}
  {2013})}\BibitemShut {NoStop}%
\bibitem [{\citenamefont {Wang}\ \emph {et~al.}(2014)\citenamefont {Wang},
  \citenamefont {Zhang}, \citenamefont {Wu},\ and\ \citenamefont
  {Zhang}}]{Wang:PhysPlasmas:2014}%
  \BibitemOpen
  \bibfield  {author} {\bibinfo {author} {\bibfnamefont {C.}~\bibnamefont
  {Wang}}, \bibinfo {author} {\bibfnamefont {{\relax Yu-J}.}~\bibnamefont
  {Zhang}}, \bibinfo {author} {\bibfnamefont {Z.-Q.}\ \bibnamefont {Wu}}, \
  and\ \bibinfo {author} {\bibfnamefont {P.}~\bibnamefont {Zhang}},\
  }\href@noop {} {\bibfield  {journal} {\bibinfo  {journal} {Phys. Plasmas}\
  }\textbf {\bibinfo {volume} {21}},\ \bibinfo {pages} {032711} (\bibinfo
  {year} {2014})}\BibitemShut {NoStop}%
\bibitem [{\citenamefont {Inogamov}\ and\ \citenamefont
  {Petrov}(2010)}]{Inogamov:JETP:2010}%
  \BibitemOpen
  \bibfield  {author} {\bibinfo {author} {\bibfnamefont {N.~A.}\ \bibnamefont
  {Inogamov}}\ and\ \bibinfo {author} {\bibfnamefont {{\relax Yu}.~V.}\
  \bibnamefont {Petrov}},\ }\href@noop {} {\bibfield  {journal} {\bibinfo
  {journal} {Journal of Experimental and Theoretical Physics}\ }\textbf
  {\bibinfo {volume} {110}},\ \bibinfo {pages} {446} (\bibinfo {year}
  {2010})}\BibitemShut {NoStop}%
\bibitem [{\citenamefont {Knyazev}\ and\ \citenamefont
  {Levashov}(2013)}]{Knyazev:COMMAT:2013}%
  \BibitemOpen
  \bibfield  {author} {\bibinfo {author} {\bibfnamefont {D.~V.}\ \bibnamefont
  {Knyazev}}\ and\ \bibinfo {author} {\bibfnamefont {P.~R.}\ \bibnamefont
  {Levashov}},\ }\href@noop {} {\bibfield  {journal} {\bibinfo  {journal}
  {Comput. Mater. Sci.}\ }\textbf {\bibinfo {volume} {79}},\ \bibinfo {pages}
  {817} (\bibinfo {year} {2013})}\BibitemShut {NoStop}%
\bibitem [{\citenamefont {Holst}\ \emph {et~al.}(2011)\citenamefont {Holst},
  \citenamefont {French},\ and\ \citenamefont {Redmer}}]{Holst:PRB:2011}%
  \BibitemOpen
  \bibfield  {author} {\bibinfo {author} {\bibfnamefont {B.}~\bibnamefont
  {Holst}}, \bibinfo {author} {\bibfnamefont {M.}~\bibnamefont {French}}, \
  and\ \bibinfo {author} {\bibfnamefont {R.}~\bibnamefont {Redmer}},\
  }\href@noop {} {\bibfield  {journal} {\bibinfo  {journal} {Phys. Rev. B}\
  }\textbf {\bibinfo {volume} {83}},\ \bibinfo {pages} {235120} (\bibinfo
  {year} {2011})}\BibitemShut {NoStop}%
\bibitem [{\citenamefont {Moseley}\ and\ \citenamefont
  {Lukes}(1978)}]{Moseley:AJP:1978}%
  \BibitemOpen
  \bibfield  {author} {\bibinfo {author} {\bibfnamefont {L.~L.}\ \bibnamefont
  {Moseley}}\ and\ \bibinfo {author} {\bibfnamefont {T.}~\bibnamefont
  {Lukes}},\ }\href@noop {} {\bibfield  {journal} {\bibinfo  {journal} {Am. J.
  Phys.}\ }\textbf {\bibinfo {volume} {46}},\ \bibinfo {pages} {676} (\bibinfo
  {year} {1978})}\BibitemShut {NoStop}%
\bibitem [{\citenamefont {Ashcroft}\ and\ \citenamefont
  {Mermin}(1976)}]{Ashcroft:1976}%
  \BibitemOpen
  \bibfield  {author} {\bibinfo {author} {\bibfnamefont {N.~W.}\ \bibnamefont
  {Ashcroft}}\ and\ \bibinfo {author} {\bibfnamefont {N.~D.}\ \bibnamefont
  {Mermin}},\ }\href@noop {} {{\selectlanguage {english}\emph {\bibinfo {title}
  {Solid State Physics}}}}\ (\bibinfo  {publisher} {Holt, Rinehart, and
  Winston},\ \bibinfo {address} {New York},\ \bibinfo {year}
  {1976})\BibitemShut {NoStop}%
\bibitem [{\citenamefont {Kresse}\ and\ \citenamefont
  {Hafner}(1993)}]{Kresse:PRB:1993}%
  \BibitemOpen
  \bibfield  {author} {\bibinfo {author} {\bibfnamefont {G.}~\bibnamefont
  {Kresse}}\ and\ \bibinfo {author} {\bibfnamefont {J.}~\bibnamefont
  {Hafner}},\ }\href@noop {} {\bibfield  {journal} {\bibinfo  {journal} {Phys.
  Rev. B}\ }\textbf {\bibinfo {volume} {47}},\ \bibinfo {pages} {558} (\bibinfo
  {year} {1993})}\BibitemShut {NoStop}%
\bibitem [{\citenamefont {Kresse}\ and\ \citenamefont
  {Hafner}(1994)}]{Kresse:PRB:1994}%
  \BibitemOpen
  \bibfield  {author} {\bibinfo {author} {\bibfnamefont {G.}~\bibnamefont
  {Kresse}}\ and\ \bibinfo {author} {\bibfnamefont {J.}~\bibnamefont
  {Hafner}},\ }\href@noop {} {\bibfield  {journal} {\bibinfo  {journal} {Phys.
  Rev. B}\ }\textbf {\bibinfo {volume} {49}},\ \bibinfo {pages} {14251}
  (\bibinfo {year} {1994})}\BibitemShut {NoStop}%
\bibitem [{\citenamefont {Kresse}\ and\ \citenamefont
  {Furthm\"uller}(1996)}]{Kresse:PRB:1996}%
  \BibitemOpen
  \bibfield  {author} {\bibinfo {author} {\bibfnamefont {G.}~\bibnamefont
  {Kresse}}\ and\ \bibinfo {author} {\bibfnamefont {J.}~\bibnamefont
  {Furthm\"uller}},\ }\href@noop {} {\bibfield  {journal} {\bibinfo  {journal}
  {Phys. Rev. B}\ }\textbf {\bibinfo {volume} {54}},\ \bibinfo {pages} {11169}
  (\bibinfo {year} {1996})}\BibitemShut {NoStop}%
\bibitem [{\citenamefont {Vanderbilt}(1990)}]{Vanderbilt:PRB:1990}%
  \BibitemOpen
  \bibfield  {author} {\bibinfo {author} {\bibfnamefont {D.}~\bibnamefont
  {Vanderbilt}},\ }\href@noop {} {\bibfield  {journal} {\bibinfo  {journal}
  {Phys. Rev. B}\ }\textbf {\bibinfo {volume} {41}},\ \bibinfo {pages} {7892}
  (\bibinfo {year} {1990})}\BibitemShut {NoStop}%
\bibitem [{\citenamefont {Povarnitsyn}\ \emph
  {et~al.}(2012{\natexlab{b}})\citenamefont {Povarnitsyn}, \citenamefont
  {Knyazev},\ and\ \citenamefont {Levashov}}]{Povarnitsyn:CPP:2012}%
  \BibitemOpen
  \bibfield  {author} {\bibinfo {author} {\bibfnamefont {M.~E.}\ \bibnamefont
  {Povarnitsyn}}, \bibinfo {author} {\bibfnamefont {D.~V.}\ \bibnamefont
  {Knyazev}}, \ and\ \bibinfo {author} {\bibfnamefont {P.~R.}\ \bibnamefont
  {Levashov}},\ }\href@noop {} {\bibfield  {journal} {\bibinfo  {journal}
  {Contrib. Plasma Phys.}\ }\textbf {\bibinfo {volume} {52}},\ \bibinfo {pages}
  {145} (\bibinfo {year} {2012}{\natexlab{b}})}\BibitemShut {NoStop}%
\bibitem [{\citenamefont {Landau}\ and\ \citenamefont
  {Lifshitz}(1980)}]{LandauLifshitz:V}%
  \BibitemOpen
  \bibfield  {author} {\bibinfo {author} {\bibfnamefont {L.~D.}\ \bibnamefont
  {Landau}}\ and\ \bibinfo {author} {\bibfnamefont {E.~M.}\ \bibnamefont
  {Lifshitz}},\ }\href@noop {} {{\selectlanguage {english}\emph {\bibinfo
  {title} {Statistical Physics. Part 1}}}},\ \bibinfo {edition} {3rd}\ ed.\
  (\bibinfo  {publisher} {Butterworth-Heinemann},\ \bibinfo {address}
  {Oxford},\ \bibinfo {year} {1980})\BibitemShut {NoStop}%
\bibitem [{\citenamefont {Anisimov}\ and\ \citenamefont
  {Rethfeld}(1997)}]{Anisimov:SPIE:1997}%
  \BibitemOpen
  \bibfield  {author} {\bibinfo {author} {\bibfnamefont {S.~I.}\ \bibnamefont
  {Anisimov}}\ and\ \bibinfo {author} {\bibfnamefont {B.}~\bibnamefont
  {Rethfeld}},\ }\href@noop {} {\bibfield  {journal} {\bibinfo  {journal}
  {Proc. SPIE}\ }\textbf {\bibinfo {volume} {3093}},\ \bibinfo {pages} {192}
  (\bibinfo {year} {1997})}\BibitemShut {NoStop}%
\bibitem [{\citenamefont {Anisimov}\ \emph {et~al.}(2006)\citenamefont
  {Anisimov}, \citenamefont {Zhakhovski\u{i}}, \citenamefont {Inogamov},
  \citenamefont {Nishihara}, \citenamefont {Petrov},\ and\ \citenamefont
  {Khokhlov}}]{Anisimov:JETP:2006}%
  \BibitemOpen
  \bibfield  {author} {\bibinfo {author} {\bibfnamefont {S.~I.}\ \bibnamefont
  {Anisimov}}, \bibinfo {author} {\bibfnamefont {V.~V.}\ \bibnamefont
  {Zhakhovski\u{i}}}, \bibinfo {author} {\bibfnamefont {N.~A.}\ \bibnamefont
  {Inogamov}}, \bibinfo {author} {\bibfnamefont {K.}~\bibnamefont {Nishihara}},
  \bibinfo {author} {\bibfnamefont {{\relax Yu}.~V.}\ \bibnamefont {Petrov}}, \
  and\ \bibinfo {author} {\bibfnamefont {V.~A.}\ \bibnamefont {Khokhlov}},\
  }\href@noop {} {\bibfield  {journal} {\bibinfo  {journal} {Journal of
  Experimental and Theoretical Physics}\ }\textbf {\bibinfo {volume} {103}},\
  \bibinfo {pages} {183} (\bibinfo {year} {2006})}\BibitemShut {NoStop}%
\bibitem [{\citenamefont {Ziman}(1960)}]{Ziman:1960}%
  \BibitemOpen
  \bibfield  {author} {\bibinfo {author} {\bibfnamefont {J.~M.}\ \bibnamefont
  {Ziman}},\ }\href@noop {} {{\selectlanguage {english}\emph {\bibinfo {title}
  {Electrons and Phonons: The Theory of Transport Phenomena in Solids}}}}\
  (\bibinfo  {publisher} {Clarendon Press},\ \bibinfo {address} {Oxford},\
  \bibinfo {year} {1960})\BibitemShut {NoStop}%
\bibitem [{\citenamefont {Apfelbaum}(2008)}]{Apfelbaum:HTHP:2008}%
  \BibitemOpen
  \bibfield  {author} {\bibinfo {author} {\bibfnamefont {E.~M.}\ \bibnamefont
  {Apfelbaum}},\ }\href@noop {} {\bibfield  {journal} {\bibinfo  {journal}
  {High Temperatures---High Pressures}\ }\textbf {\bibinfo {volume} {37}},\
  \bibinfo {pages} {253} (\bibinfo {year} {2008})}\BibitemShut {NoStop}%
\bibitem [{\citenamefont {Desjarlais}(2001)}]{Desjarlais:CPP:2001}%
  \BibitemOpen
  \bibfield  {author} {\bibinfo {author} {\bibfnamefont {M.~P.}\ \bibnamefont
  {Desjarlais}},\ }\href@noop {} {\bibfield  {journal} {\bibinfo  {journal}
  {Contrib. Plasma Phys.}\ }\textbf {\bibinfo {volume} {41}},\ \bibinfo {pages}
  {267} (\bibinfo {year} {2001})}\BibitemShut {NoStop}%
\bibitem [{\citenamefont {Redmer}(1999)}]{Redmer:PRE:1999}%
  \BibitemOpen
  \bibfield  {author} {\bibinfo {author} {\bibfnamefont {R.}~\bibnamefont
  {Redmer}},\ }\href@noop {} {\bibfield  {journal} {\bibinfo  {journal} {Phys.
  Rev. E}\ }\textbf {\bibinfo {volume} {59}},\ \bibinfo {pages} {1073}
  (\bibinfo {year} {1999})}\BibitemShut {NoStop}%
\bibitem [{\citenamefont {Kuhlbrodt}\ and\ \citenamefont
  {Redmer}(2000)}]{Kuhlbrodt:PRE:2000}%
  \BibitemOpen
  \bibfield  {author} {\bibinfo {author} {\bibfnamefont {S.}~\bibnamefont
  {Kuhlbrodt}}\ and\ \bibinfo {author} {\bibfnamefont {R.}~\bibnamefont
  {Redmer}},\ }\href@noop {} {\bibfield  {journal} {\bibinfo  {journal} {Phys.
  Rev. E}\ }\textbf {\bibinfo {volume} {62}},\ \bibinfo {pages} {7191}
  (\bibinfo {year} {2000})}\BibitemShut {NoStop}%
\end{thebibliography}
%\bibliographystyle{apsrev4-1}

\end{document}